\documentclass[pre,twocolumn,showpacs,a4paper,10pt]{revtex4-1}

\usepackage{amsfonts,amssymb,amsmath,amsthm,latexsym,times}
\usepackage{graphicx, color}
\usepackage[english]{babel}
\usepackage{float}
\usepackage{enumitem}
\usepackage{siunitx}

\usepackage{cancel}
\usepackage{amssymb,scalerel}

\def\ci{\perp\!\!\!\perp}

\newcounter{thm}
\newcounter{ex}
\newcounter{re}
\newtheorem{Theorem}[thm]{Theorem}

\begin{document}
\title{Quantifying information transfer and mediation along causal pathways in complex systems}

\author{Jakob Runge}
\affiliation{ Potsdam Institute for Climate Impact Research, P.O. Box 60 12 03, 14412 Potsdam, Germany \\
Department of Physics, Humboldt University, Newtonstr. 15, 12489 Berlin, Germany
        }
\date{\today}

\begin{abstract}
Measures of information transfer have become a popular approach to analyze interactions in complex systems such as the Earth or the human brain from measured time series. Recent work has focused on causal definitions of information transfer aimed at decompositions of predictive information about a target variable, while excluding effects of common drivers and indirect influences. While common drivers clearly constitute a spurious causality, the aim of the present article is to develop measures quantifying different notions of the strength of information transfer along indirect causal paths, based on first reconstructing the multivariate causal network (\emph{Tigramite} approach). Another class of novel measures quantifies to what extent different intermediate processes on causal paths contribute to an interaction mechanism to determine pathways of causal information transfer. The proposed framework complements predictive decomposition schemes by focusing more on the interaction mechanism between multiple processes.
A rigorous mathematical framework allows for a clear information-theoretic interpretation that can also be related to the underlying dynamics as proven for certain classes of processes. Generally, however, estimates of information transfer remain hard to interpret for nonlinearly intertwined complex systems. But, if experiments or mathematical models are not available, measuring pathways of information transfer within the causal dependency structure allows at least for an abstraction of the dynamics. 
The measures are illustrated on a climatological example to disentangle pathways of atmospheric flow over Europe.
\end{abstract}

\maketitle

%%%%%%%%%%%%%%%%%%%%%%%%%%%%%%%%%%%%%%%%%%

\section{Introduction}
\label{sec:source_entropy}

The availability of vast amounts of time series data from such complex systems as the Earth or the human brain and body has given rise to a plethora of time series analysis methods aimed at understanding interactions between regions or subprocesses in these complex systems. 
Of a particular interest are methods to quantify some notion of \emph{information flow} or \emph{information transfer} within the complex system. In neuroscience \cite{Bullmore2009} and climate research \cite{Tsonis2008,Donges2009}, such interpretations have often been based on pure pairwise correlation analyses. But towards measuring information transfer, the method should, firstly, be general enough to include also nonlinear associations. This can be achieved in an information-theoretic framework with measures such as mutual information (MI) \cite{Cover2006}.
Secondly, networks reconstructed from pairwise measures of association (be it cross-correlation or MI) do not allow to assess the propagation of information or hypothetical perturbations in a causal sense: For example, an interaction like $X\leftarrow Z \rightarrow Y$ would imply that $X$ and $Y$ are correlated even though no perturbations originating in $X$ can actually reach $Y$, or vice versa.

An important step towards deeper insights has, therefore, been achieved by methods that are capable of inferring a statistical notion of directionality or even causal interactions which have been applied to the climate system \cite{Hlinka2013,Deng2014a,Schleussner2013,Balasis2013,Zerenner2014,Runge2014c}, the human brain \cite{Niso2013,Wibral2014,Lehnertz2015}, and to disentangle cardiovascular processes \cite{Faes2011,Faes2014,Runge2014e}, among others.
Causal associations between subprocesses can be visualized as links in a complex interaction network. 
A full causal reconstruction of a link $X\to Y$ can only be achieved under the in most cases unrealistic assumption that all possible other influences on $X$ and $Y$ can be included in the analysis \cite{Granger1969,Spirtes2000}, or if the system can be experimentally manipulated within Pearl's causal effect framework \cite{Pearl2000}. Usually, it is impossible to exclude all other influences and large complex systems can typically not be easily experimentally manipulated. Causal inference based on data-analysis methods, therefore, provides only a first step and the term ``causal'' can then only be understood to be meant relative to the system under study, i.e., the processes that comprise the nodes of the network.

Two tasks need to be addressed to measure a causal notion of information transfer from time series of complex systems:
\begin{enumerate} 
\item Reconstructing the causal network,
\item Quantifying causal information transfer.
\end{enumerate} 
In this article we will focus on the quantification part, the reconstruction problem has been addressed by the author in Ref.~\cite{Runge2012prl}.
As further reviewed below, previous works have mainly considered a decomposition of the predictive information in direct drivers of a process $Y$. In the present article, we ask a different question: How does information \emph{originating} in a process $X$ propagate also on \emph{indirect} paths through the causal interaction network? How strong is it and which intermediate processes on causal pathways are contributing to such a mechanism? 

The paper is organized as follows: In the remainder of this introductory section, we review recent approaches to measuring information transfer in complex systems and sketch the basic idea underlying the present approach. In Sect.~\ref{sec:info_theory} we recall basic concepts of information theory and in Sect.~\ref{sec:tsg} introduce the concept of \emph{time series graphs} as the causal basis of the present approach. In Sect.~\ref{sec:tigramite} we introduce the novel measures based on time series graphs to quantify interactions along paths and mediation and distinguish them from transfer entropy-related approaches. In Sect.~\ref{sec:analytical_examples} we extensively analyze the measures with analytical and numerical examples and provide theorems that foster a more rigorous mathematical and dynamical understanding to facilitate the interpretability of the proposed measures. Section~\ref{sec:discussion} discusses the theoretical results and relation to linear measures of causal effect in Pearl's framework \cite{Pearl2013}, and gives an outlook to applications of the novel measures in complex network theory.
Finally, Sect.~\ref{sec:climate} gives an illustrative application to climatological time series and Sect.~\ref{sec:conclusions} concludes the paper. The appendix contains proofs of the theorems.

\subsection{Quantifying causal information transfer}
\label{sec:dependence_axioms}

Compared to the first task of detecting causal interactions, more or less a binary question, the second task of quantifying causal information transfer is much more ambiguous to define in a universal way which has led Smirnov \cite{Smirnov2014,Smirnov2015} to question the goal of assessing a ``causal coupling strength'' and instead measure ``how the coupling manifests itself in the dynamics'' in an \emph{interventional-effect} causal framework as proposed by Pearl \cite{Pearl2000}. In Ref.~\cite{Lizier2010} the term `information transfer' is even distinguished from `information flow' where the latter is meant in a causal sense based on interventions. This framework, however, necessitates either to experimentally manipulate the system, or to have a mathematical model to perform ``virtual interventions''. 
To some extent causal effects can also be extracted if the time series cover the whole state space or attractor of the complex system \cite{Smirnov2014} such that virtual interventions can be drawn `randomly' from the stationary distribution. 
In a mathematical model the strength of a coupling mechanism can often be related to model coefficients and a plethora of methods exists that implement the model-based concept of Granger causality \cite{Granger1969}. These range from classical linear autoregressive models in the form of the \emph{directed transfer function} \cite{Kaminski2001,Korzeniewska2003,Blinowska2006}, to slightly less restrictive approaches such as \emph{partial directed coherence} using spectral estimators  \cite{Baccala2001,Schelter2006c,Jachan2009,Sommerlade2009,Schelter2009}, \emph{extended Granger causality} with local linear embeddings in phase space \cite{Chen2004}, or kernel estimators \cite{Marinazzo2008}, to name just a few. All these approaches still involve strong assumptions about the dependencies and share the problem that the model might be misspecified. This implies that the model may not adequately represent important interactions such as the complicated interplay between El Ni\~no Southern Oscillation and the Indian Monsoon in the climate system \cite{Stocker2013} or neural interactions where even a fully physical model is lacking.
% Information-theoretic approaches have the appeal that they capture almost any form of statistical association. 
% Generally, however, especially in large spatio-temporal systems such as the Earth or the human brain, mathematical models are either not available or computationally expensive and may not adequately represent important interactions such as the complicated interplay between El Ni\~no Southern Oscillation and the Indian Monsoon in the climate system \cite{Stocker2013} or neural interactions where even a fully physical model is lacking.

If it is not possible to measure ``how the coupling manifests itself in the dynamics'', information-theoretic quantifiers can at least help to measure ``how the causal coupling manifests itself in the exchange of entropy between the subprocesses'' in an information-theoretic framework capturing almost any form of statistical association. Here `causal' is meant relative to the observed process as discussed above. This approach aims to distinguish different contributions based on the Markovian conditional independence structure of the multivariate process as an abstraction of the dynamics.

There are few works considering multivariate definitions of information transfer and their interpretation.
In Ref.~\cite{Faes2015a}, the central concept is to decompose the predictive information about the next time step of a subprocess $Y$ into the MI between $Y$ and its own past as the \emph{information storage}, the \emph{partial transfer entropy} from another subprocess $X$, and the TE between $Y$ and the remaining process.
In Refs.~\cite{Stramaglia2012,Stramaglia2014} another decomposition is proposed to detect redundant and synergetic contributions of driving variables.
Liang \cite{Liang2014,Liang2015a} presents a rigorous approach based on the underlying Langevin description of a system to define the contributions of internal and external driving to the evolution of the entropy of a subprocess $Y$. This approach is, however, based on the knowledge of the deterministic-stochastic equations of the system, but in principle it can also be estimated from time series alone involving numerical optimization problems.
In Refs.~\cite{ay2008information,Janzing2013} an idea is described that is similar to the present approach in that there the question of quantifying the strength of links is seen as a second step based on the known causal network.  Ay \emph{et al.} \cite{ay2008information} address the problem from an interventionalist perspective using Pearl's do-calculus \cite{Pearl2000} which we do not further discuss here since we assume the process to be not manipulable. Janzing \emph{et al.} \cite{Janzing2013} define the strength of a link $X\to Y$ by considering the thought experiment of an attacker `cutting the link' and feeding in the distribution of $X$ as an input, arriving at a measure that is not a conditional mutual information anymore, which we use here to measure the transfer of information. Also, the authors state that it is difficult to quantify also indirect effects in their framework.
In general, there are different ways to define measures and different research questions demand different properties.

\subsection{The idea of momentary information} \label{sec:motivation}
%%%%%%%%%%%%%%%%%%%%%%%%%%%%%%%%%%%%%%%%%%%%%%%%%%%%%%%%%%%%%%%%%%%%%%%%%%%%%%%%%%%%%%%%%%%%%%
\begin{figure}[t!]
\begin{center}
\includegraphics[width=.8\columnwidth]{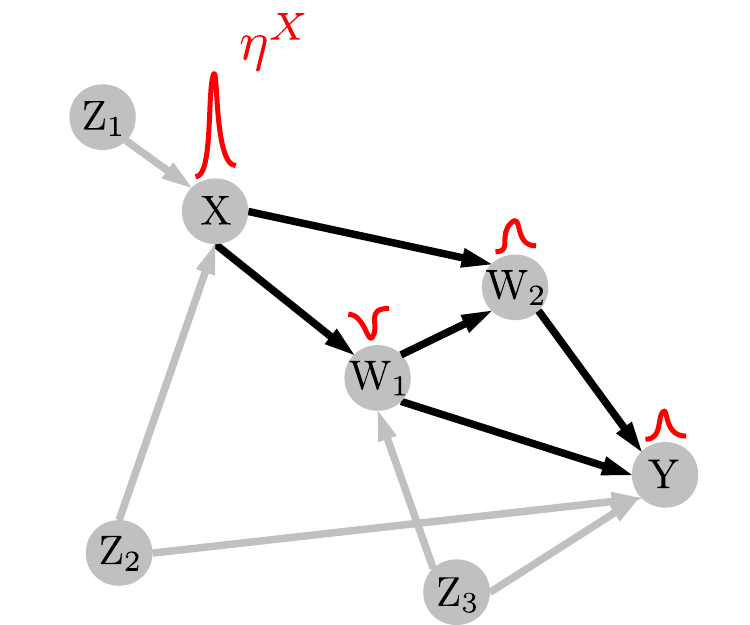}
\end{center}
\caption[]{(Color online) Consider a realization of dynamical noise $\eta^X$ driving subprocess $X$ as a perturbation. Coupling mechanisms along different causal paths (black lines) transform such a perturbation, and the total effect on $Y$ some time later can also depend on how intermediate processes nonlinearly interact with each other as shown in Sect.~\ref{sec:interaction_examples}. The central idea of the momentary information transfer measures presented in this article is to information-theoretically quantify the general effect of such perturbations and isolate it from common drivers in the past such as $Z_2$, but also $Z_1$ and the past of $X$. To also quantify how much intermediate processes such as $(W_1,\,W_2)$ on causal paths mediate information, it will also be important to exclude common drivers like $Z_3$.
}
\label{fig:mitp_motivation}
\end{figure}

The approach to measures of causal information transfer formally introduced in Sect.~\ref{sec:tigramite} is based on the fundamental concept of \emph{source entropy}, also termed the \emph{entropy rate} \cite{Shannon1948,Shannon1963}, and was introduced for the special case of bivariate ordinal pattern time series in Ref.~\cite{Pompe2011}. 
Consider a symbol-generating process $X$. At each time $t$ a realization $x_t$ is generated. Now the source entropy of $X_t$ measures the uncertainty about $x_t$ before its observation if all former observations $(x_{t-1},\, x_{t-2},\ldots)$ are known (entropies will be formally introduced in Sect.~\ref{sec:info_theory}). For a completely deterministic non-chaotic system the source entropy will always be zero, but for a real world process there will always be some uncertainty stemming from \emph{dynamical noise}. This type of noise is to be distinguished from \emph{observational noise} which usually contaminates each measured time series \cite{Schreiber1995}, but has no effect on the dynamics of the process. Dynamical noise might occur due to unresolved smaller-scale processes and can be modeled by including a random variable in the system. More formally, consider a subprocess $X$ of a multivariate process $\mathbf{X}$ with infinite past $\mathbf{X}_t^-=(\mathbf{X}_{t-1},\,\mathbf{X}_{t-2},\,\ldots)$, that is described by the discrete-time equation

\begin{align} \label{eq:source_process}
X_t = f\left(Z_{1,t-\tau_1},\,Z_{2,t-\tau_2},\,\ldots, \eta^X_t\right),
\end{align}

with some arbitrary function $f$ of other subprocesses at past times $Z_{1,t-\tau_1}$, $Z_{2,t-\tau_2}$, $\ldots$ $\in \mathbf{X}_t^-$  and the random part subsumed under $\eta^X_t$. The uncertainty of an outcome $x_t$ will \emph{on average} be reduced if a realization of the past $Z_{1,t-\tau_1},\,Z_{2,t-\tau_2},\,\ldots$ is known. But for non-zero $\eta^X_t$ there will always be some ``surprise'' left when observing $x_t$. This surprise gives us information and the expected information here is the source entropy $H(X_t|\mathbf{X}_t^-)$ of $X$. If the dynamical noise $\eta^X_t$ occurs additively in Eq.~(\ref{eq:source_process}), then $H(X_t|\mathbf{X}_t^-)=H(\eta^X_t)$.
Due to measurement errors or observational noise $\epsilon$, we will in general not be able to estimate the source entropy alone, but only $H(X_t+\epsilon^X_t|\mathbf{X}_t^-+\epsilon^{\mathbf{X}^-}_t)$.
Even assuming a perfect measurement apparatus for a deterministic dynamical system without dynamical noise, the entropy rate $h^{\rm symb}$ -- since it is computed by creating a symbol sequence from a coarse graining in phase-space -- depends on some resolution parameter $r$. Then the limit $\lim_{r\to 0}h^{\rm symb}$ might exist and is called the \emph{Kolmogorov-Sinai entropy}. If this limit is finite and larger than zero, the system is called chaotic.
But here we study stochastic, discrete time processes because the finite set of measured variables of a complex system like the Earth will never perfectly describe the full system's state and all remaining processes contribute to dynamical noise (implying that the Kolmogorov-Sinai entropy diverges).

While the focus in Refs.~\cite{Faes2015a,Stramaglia2012} and related works is on decompositions of predictive information on the basis of transfer entropy as an information-theoretic generalization of Granger causality, the concept here is more similar to \emph{Sims} causality, see, e.g., \cite{Florens1982}, which takes into account not only direct, but also indirect causal effects. Sims causality is based on measuring to what extent $X$ at time $t$ helps in predicting $Y$ at times $t'>t$ in the future excluding the past of $X$ and also the present of all other processes, i.e., $\mathbf{X}_{t+1}^-=(\mathbf{X}_{t},\,\mathbf{X}_{t-1},\,\ldots)$. 
In model~(\ref{eq:source_process}) excluding the past essentially isolates the dynamical noise $\eta^X_t$ and our goal is now to quantify the information transfer emanating from $\eta^X_t$ into the future (Fig.~\ref{fig:mitp_motivation}).

With this central idea we define two pairs of measures for two purposes: (1)~to quantify the information transfer between two causally linked processes and along causal paths and (2)~the mediation of intermediate processes. For each of these tasks we define two measures quantifying different notions of information transfer: Both have in common the above idea to extract information originating in process $X$ only at the lagged time $t-\tau$ and are conditioned in order to measure only information transfer along causal paths. These measures, thus, complement alternative decomposition approaches such as in Refs.~\cite{Faes2015a,Liang2014,Stramaglia2012}.
The second measure further attempts to exclude the influence of other drivers of $Y$ or intermediate path nodes to isolate the whole causal information pathway and fulfill a generalized property of \emph{coupling strength autonomy} as proposed in previous work \cite{Runge2012b}. In the present context the property of coupling strength autonomy demands that the measure should be uniquely determined by the interaction of the two processes, $X,\,Y$ in the previous example and possibly intermediate other processes $W$, alone and in a way autonomous of how these are driven by the remaining processes. To understand this, consider a simple example: Suppose we have two interacting processes $X$ and $Y$ and a third process $Z$, that drives both of them. Then a bivariate measure of coupling strength between $X$ and $Y$ such as MI will be influenced by the common input of $Z$, while our demand is, that the measure should be autonomous of the interactions of $X$ and $Y$ with $Z$. 
% The concept of \emph{momentary information} originating in a process $X$ implements this property. 

In summary, this paper generalizes the idea underlying Ref.~\cite{Runge2012b} to use the reconstructed causal network for quantifying general causal interactions. This framework is called the \emph{Tigramite} approach (\emph{\underline{Ti}me series \underline{gra}ph based \underline{M}easures of \underline{I}nformation \underline{T}ransf\underline{e}r}), which is also the abbreviation of the accompanying software package (available on the author's website). Table~\ref{tab:tigramite} gives an overview over different ways to use the time series graph for defining causal information transfer measures.

Pearl \cite{Pearl2000} defines the causal effect of $X$ on $Y$ by the hypothetical intervention of \emph{experimentally setting} a variable $X$ to a certain value $x$. Then the \emph{post-interventional} distribution $P(Y=y~|~do(X=x))$, which involves the $do$-operator and is not the same as the conditional distribution, is used to assess whether and in what way $X$ affects $Y$. As mentioned before, however, we assume a non-manipulable complex system and, therefore, study a weaker notion of causality. From observational data alone, causal effects can only be estimated (or \emph{identified}) under certain assumptions about the underlying process and the kind of interventions \cite{Eichler2010a,Pearl2000}. In Sect.~\ref{sec:causal_effect} we discuss Pearl's causal effect for linear models.

%%%%%%%%%%%%%%%%%%%%%%%%%%%%%%%%%%%%%%%%%%

\section{Information-theoretic preliminaries} \label{sec:info_theory}

\subsection{Conditional mutual information}
\label{sec:conditional_mutual_information}
%%%%%%%%%%%%%%%%%%%%%%%%%%%%%%%%%
\begin{figure}[t!]
\begin{center}
\includegraphics[width=1.\columnwidth]{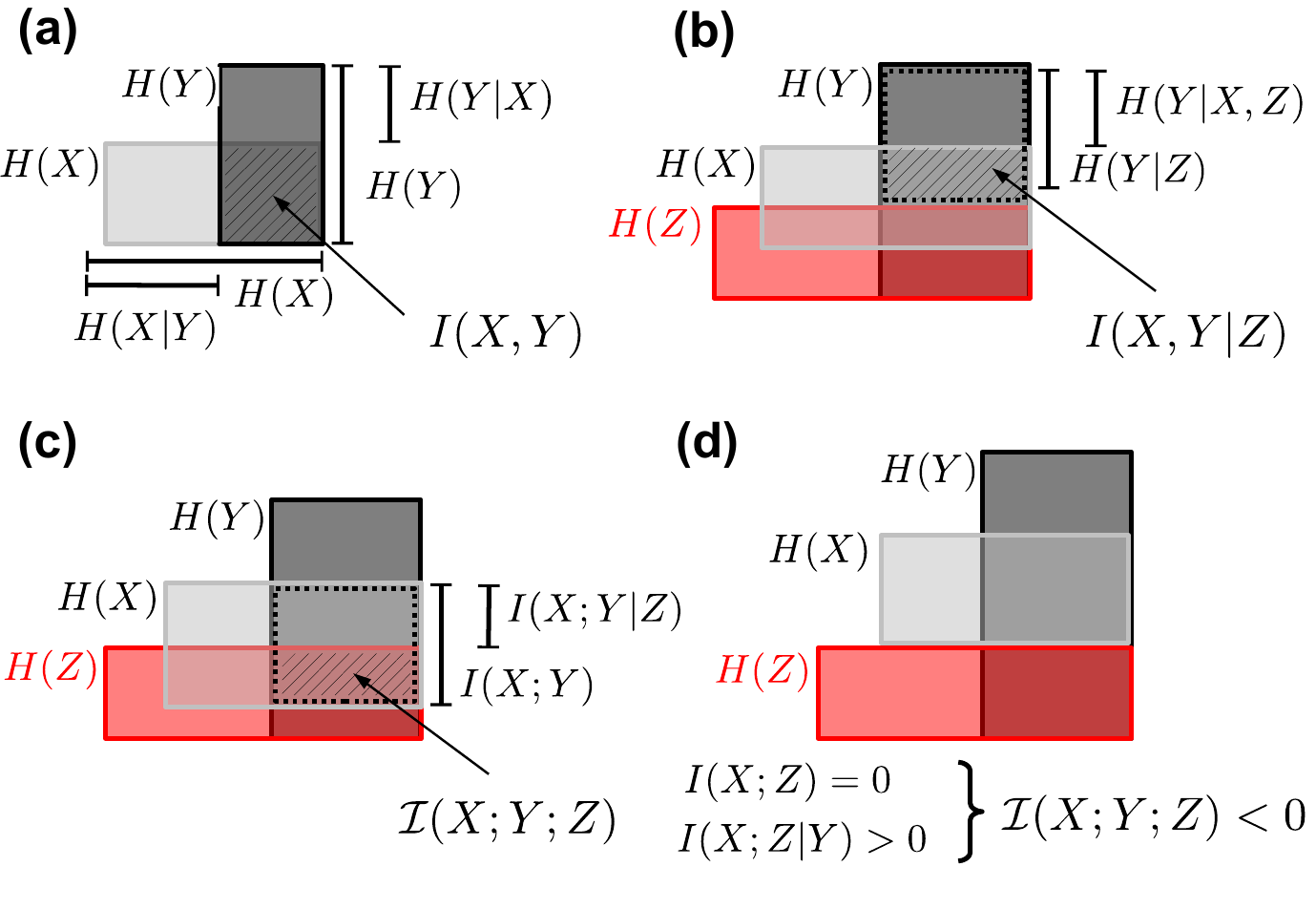}
\end{center}
\caption[Venn diagrams of (conditional) mutual information and interaction information.]{(Color online) Venn diagrams of (a) mutual information, (b) conditional mutual information, (c) positive interaction information, and (d) negative interaction information. The latter case, where  the entropies of $X$ and $Z$ do not `overlap' anymore, demonstrates that the analogy between entropies and sets should not be overinterpreted.}
\label{fig:info_theory}
\end{figure}
%%%%%%%%%%%%%%%%%%%%%%%%%%%%%%%%%%%%

The most important information-theoretic measure on which the quantities discussed in this article are based is the \emph{conditional mutual information} (CMI) given by

\begin{align} 
& I(X;Y | Z ) \nonumber\\
&= H(Y|Z) - H(Y|X,Z) = H(X|Z) - H(X|Y,Z) \label{eq:def_cmi_diff}  \\
&=\int p(z)  \iint  p(x,y|z) \log \frac{ p(x,y |z)}{p(x|z)\cdot p(y |z)} \,dx dy dz\,,  \label{eq:def_cmi}
\end{align}

with Shannon's entropy $H$ \cite{Shannon1948,Shannon1963} as a measure of the uncertainty about outcomes of a process. Mutual information (MI), on the other hand is a measure of the reduction of this uncertainty if another process is measured and CMI can be phrased as the MI between $X$ and $Y$ that is not contained in a third variable $Z$.  Here we use the natural logarithm to measure CMI and derived measures in $nats$. Note that $X$, $Y$, and $Z$ can also be vectors. Just like MI, CMI is non-negative (which can be shown using Jensen's inequality \cite{Cover2006} and holds for the continuous as well as the discrete case) and symmetric in its first two arguments $I(X;Y|Z) = I(Y;X|Z)$. 
Further, according to Eq.~(\ref{eq:def_cmi}), CMI measures the Kullback-Leibler distance \cite{Kullback1951,Cover2006} between the distributions $p(x,y|z)$ and the distribution for the independent case $p(x|z)p(y|z)$ and is zero if and only if $X$ and $Y$ are independent \emph{conditionally on $Z$}.
% \begin{align} \label{eq:cond_indep_cmi}
% X \ci Y | Z ~~~&\Longleftrightarrow~~~ p(x,y|z) = p(x|z) p(y|z)   \nonumber\\
% 				&\Longleftrightarrow~~~ I(X;Y|Z)=0.
% \end{align}
This property makes CMI especially useful to measure conditional independence as needed in the definition and estimation of causal graphs (Sect.~\ref{sec:tsg}). Figures~\ref{fig:info_theory}(a) and (b) visualize MI and CMI in Venn diagrams as a difference of conditional entropies. In this representation also the symmetry in the arguments is obvious. 

\subsection{Interaction information}

Just like MI and CMI are differences of conditional entropies, also the difference of CMIs has an interesting interpretation that we will utilize to measure the effect of one random variable on the interaction between two others. Such a measure has been studied in Refs.~\cite{Abramson1963,Tsujishita1995,Leydesdorff2009} under the name \emph{multiple information}. We use the term \emph{interaction information} with the symbol $\mathcal{I}$, which is symmetrically defined as

\begin{align} \label{eq:def_interaction_information}
\mathcal{I}(X;Y;Z) &= I(X;Y) - I(X;Y|Z) \\
                    &= I(Y;Z) - I(Y;Z|X) \nonumber\\
                    &= I(Z;X) - I(Z;X|Y) \nonumber.
\end{align}

In Refs.~\cite{McGill1954,Jakulin2003} this quantity is defined with the signs reversed, but the above definition is more consistent with the definition of CMI in Eq.~(\ref{eq:def_cmi_diff}). 
It is also straightforward to define the \emph{conditional interaction information}

\begin{align}
\mathcal{I}(X;Y;Z|W) &= I(X;Y|W) - I(X;Y|Z,W).
\end{align}

Contrary to CMI, the (conditional) interaction information can also be negative and is bounded by

\begin{align} \label{eq:interaction_bounds}
&-\min(I(X;Y|Z,W),I(Y;Z|X,W),I(Z;X|Y,W)) \nonumber\\
&\leq  \mathcal{I}(X;Y;Z|W) \nonumber\\
&\leq \min(I(X;Y|W),I(Y;Z|W),I(Z;X|W)).
\end{align}

The possible negativity also shows that the visualization in Fig.~\ref{fig:info_theory}(c) as sets in Venn diagrams should not be overinterpreted. In Fig.~\ref{fig:info_theory}(d) a case is shown where $X$ and $Z$ are \emph{unconditionally} independent, but conditionally dependent leading to $I(X;Z|Y) \geq I(X;Z)$ and, therefore, a negative interaction information. 
That this property can actually by intuitively understood will be studied in examples in Sect.~\ref{sec:analytical_examples}. 

\subsection{Estimation of (conditional) mutual information} \label{sec:cmi_estimation}

In the examples and applications we use a nearest-neighbor estimator \cite{Kraskov2004a,Frenzel2007} that is most suitable for variables taking on a continuous range of values and has much less bias than the commonly used binning estimators. This estimator has as a free parameter the number of nearest-neighbors $k$ which determines the size of hyper-cubes around each (high-dimensional) sample point. Small values of $k$ lead to a lower estimation bias but higher variance and vice versa. For independence tests, a higher $k$ with lower variance is more important while for estimates of the CMI value a smaller $k$ is recommended. 
% Ref.~\cite{Runge2014d} contains some numerical studies on the statistical power as an independence test and the estimator's bias. 
Note that for an estimation from (multivariate) time series stationarity is required.

%%
%%
%%
% \clearpage
\section{Time series graphs and causal paths} \label{sec:tsg}

The here proposed framework to use the reconstructed causal network for quantifying general causal interactions (\emph{Tigramite} approach) is based on the concept of time series graphs and causal paths as defined in the following.

\subsection{Time series graphs}

A \emph{time series graph} \cite{Dahlhaus2000,Eichler2011} is a certain type of graphical model \cite{lauritzen1996graphical} for the case of time-ordered data and visualizes the Markovian conditional independence properties of a multivariate time-dependent process, i.e., how the joint density of the multivariate process $\mathbf{X}$ (including its lags) factorizes. Figures~\ref{fig:path_measures}(a,b) show examples.
Each node in a time series graph represents a subprocess of a multivariate discrete time process $\mathbf{X}$ at a certain time $t$. Directed links between subprocesses (or nodes) $X_{t-\tau}$ and $Y_t$ for $\tau>0$ are marked by an arrow and defined by

\begin{align}  \label{eq:def_graph}
  X_{t-\tau}~\to~Y_t ~\Longleftrightarrow~I(X_{t-\tau}; Y_t| \mathbf{X}_t^-\setminus \{X_{t-\tau}\}) > 0\,,
\end{align}

with infinite past $\mathbf{X}_t^-=(\mathbf{X}_{t-1},\,\mathbf{X}_{t-2},\,\ldots)$, i.e., if they are not independent conditionally on the past of the whole process, which implies a lag-specific Granger causality with respect to $\mathbf{X}$.
If $Y\neq X$ we say that the link $X_{t-\tau} \to Y_t$ represents a \textit{coupling or cross-link at lag} $\tau$, while for $Y=X$ it represents an \textit{autodependency or auto-link at lag} $\tau$. 

Since often also contemporaneous associations are of interest, we also define links between $X_t$ and $Y_t$ as in previous works \cite{Runge2012prl,Runge2012b} by

\begin{align} \label{eq:def_graph_contemp}
X_t{\--} Y_t ~\Longleftrightarrow~I(X_t; Y_t\,|\, \mathbf{X}_{t+1}^- {\setminus} \{X_t,Y_t\}) > 0\,,
\end{align}

where also the contemporaneous present $\mathbf{X}_t{\setminus}\{X_t,Y_t\}$ is included in the condition. Note that stationarity implies that $X_{t-\tau} \to Y_t$ whenever $X_{t'-\tau}\to Y_{t'}$ for any $t'$ and correspondingly for contemporaneous links. In Ref.~\cite{Eichler2011} also another version of contemporaneous links is defined, marked by a dashed line:

\begin{align} \label{eq:def_graph_contemp_dashed}
% I(X_t; Y_t\,|\, \mathbf{X}_{t+1}^- {\setminus} \{X_t,Y_t\}) > 0\,,
X_t~\text{- - -}~Y_t ~\Longleftrightarrow~ I(X_t; Y_t\,|\, \mathbf{X}_{t}^- ) > 0\,.
\end{align}

In the case of a multivariate autoregressive process, the latter definition corresponds to non-zero entries in the covariance matrix of the innovations, while the former corresponds to non-zero entries in the  \emph{inverse} covariance matrix \cite{Eichler2011}.
One problem with Definition~(\ref{eq:def_graph_contemp}) is that it can potentially cause spurious links if, e.g., $X_t$ and $Y_t$ are independent (also of the past), but both causally drive another process $Z_t$ instantaneously, i.e., at the same time $t$, which might not be resolved due to a too coarse time sampling interval. Then $I(X_t; Y_t) = 0$, but $I(X_t; Y_t~|~Z_t) > 0$ due to the `conditioning on a common child' effect, see e.g. \cite{Chicharro2014}, which is shown in Fig.~\ref{fig:info_theory}(d). In this work, we are not considering instantaneous causal effects, but to circumvent this problem in practice, one can consider contemporaneous effects only if both Definitions (\ref{eq:def_graph_contemp_dashed}) and (\ref{eq:def_graph_contemp}) are satisfied.
Note that both definitions result in slight differences in the definition of open and blocked paths through contemporaneous links as discussed further below.

In Refs.~\cite{Runge2012prl,Runge2014a} a consistent algorithm for the estimation of the above-defined time series graphs by iteratively inferring the parents and, in a second step, also the neighbors is discussed. This challenging problem is not further addressed here and involves demands such as consistency (i.e., that the algorithm converges to the true graph for infinite sample sizes), statistical power, underlying assumptions (e.g., faithfulness \cite{Spirtes2000}), or computational complexity (partly addressed in Ref.~\cite{Runge2014b}).

\subsection{Causal paths} \label{sec:tsg_paths}

The measures introduced in Sect.~\ref{sec:tigramite} are CMIs based on paths and different sets of conditions which we determine from the sets of parents and neighbors of a node $Y_t$ defined, respectively, as

\begin{align}
\mathcal{P}_{Y_t} &= \{Z_{t-\tau}:~ Z\in \mathbf{X},~\tau>0,~Z_{t-\tau}\to Y_t\}\,,\\
\mathcal{N}_{Y_t} &= \{X_t:X\in\mathbf{X}, X_t {\--} Y_t\}.
\end{align}

Our main interest lies in \emph{causal paths} in the time series graph which are defined as directed paths, i.e., containing only motifs $\to\,\bullet\,\to$ (assuming that the arrow of time in the time series graph goes to the right). But there are also other paths on which information is shared even though no causal interventions could `travel' along these. In general \cite{Eichler2011}, in the above defined time series graph with solid contemporaneous links a path between two nodes $u$ and $v$ is called \emph{open} if it contains only the motifs $\to\,\bullet\,\to$, $\leftarrow\,\bullet\,\to$, ${\--}\,\bullet\,\to$, or ${\--}\,\bullet\,{\--}$. On the other hand, if any motif on a path is $\to\,\bullet\,\leftarrow$ or $\to\bullet\,{\--}$, the path is blocked. Nodes in such motifs are also called \emph{colliders}. If we now consider a \emph{separating or conditioning set $\mathcal{S}$}, openness and blockedness of conditioned motifs reverse, i.e., denoting a conditioned node by $\blacksquare$, the motifs $\to\,\blacksquare\,\to$, $\leftarrow\,\blacksquare\,\to$, $\,{\--}\,\blacksquare\,\to$, and ${\--}\,\blacksquare\,{\--}$ are blocked and the motifs $\to\,\blacksquare\,\leftarrow$ and $\to\blacksquare\,{\--}$  become open. Note that for the alternative definition of contemporaneous links Eq.~(\ref{eq:def_graph_contemp_dashed}) marked with dashed lines, the motif $\,\text{- - -}\,\bullet\,\text{- - -}$ is blocked while the conditioned motif $\,\text{- - -}\,\blacksquare\,\text{- - -}$ is open.

Two nodes $u$ and $v$ are \emph{separated given a set $\mathcal{S}$} if all paths between the two are blocked. Conversely, two nodes are \emph{connected given a set $\mathcal{S}$} if at least one path between the two is open.  The Markov property, which we assume throughout, now relates separation in the time series graph to conditional independence relations in the underlying process which can be quantified with CMI (as a conditional independence measure):

\begin{align} \label{eq:markovity}
\text{$u$ and $v$ separated given $\mathcal{S}$} ~\implies~ I(u;v~|~\mathcal{S}) = 0\,.
\end{align}

\begin{table*}[t!]
\centering
\begin{tabular}{l|p{0.22\linewidth}|p{0.22\linewidth}|p{0.3\linewidth}}  
% \hline
% \hline
 & Granger causality / TE - type & Sims causality - type & Causal information pathways - type  \\ 
\hline
Conditioned on & parents of target process $Y$ & parents and neighbors of source process $X$ & parents and neighbors of source and parents of all pathway variables \\
% \hline
Transfer measures & (D)TE (not lag-specific), ITY   & ITX  &  MIT (causal links only), MITP \\
% \hline
Interaction measures & $-$ & IIX   & MII \\ 
\hline 
\end{tabular}
\caption{Three different types of time series graph-based measures of information transfer (Tigramite approach). Transfer measures refer to CMI-based quantities to measure information transfer between two variables, interaction measures to the interaction information-based quantities between multiple variables. Decomposed transfer entropy (DTE) was introduced in Ref.~\cite{Runge2012prl}.}
\label{tab:tigramite}
\end{table*}

The path-based CMIs are constructed with conditions to block all non-causal paths and only leave open causal paths. In particular, also \emph{contemporaneous sidepaths}, which start with one or more contemporaneous links followed by a directed path $u\,{\--}\bullet\,{\--}\,\cdots\,{\--}\,\bullet\,\to\,\cdots\,\to v$, need to be blocked.
Note that we do not consider contemporaneous causal effects here which might occur due to a too low sampling rate of the process.

\section{Time series graph based measures of information transfer (TiGraMITe approach)} \label{sec:tigramite}

In the following we briefly discuss the transfer entropy ansatz to measuring information transfer and introduce our novel approach to quantify different aspects of information transfer through causal links and paths. Table~\ref{tab:tigramite} provides an overview over these different classes of measures.
As mentioned in the introduction, the proposed measures of information transfer are CMIs based on different sets of conditions which we determine from the reconstructed time series graph. The \emph{Tigramite} approach has the advantage of a low-dimensional estimation problem without arbitrary truncation parameters like in the original definition of transfer entropy involving infinite vectors.

\subsection{Transfer entropy ansatz}
\label{sec:ity}
% \begin{figure}[t!]
% %\epsfig{file=fig1.eps} 
% %\includegraphics[width=\columnwidth]{entropy_flow.pdf}
% \begin{center}
% %\includegraphics[width=.6\columnwidth]{images/pre_fig2_v2.pdf}
% \includegraphics[width=1.\columnwidth]{mit_ity.pdf}
% \end{center}
% \caption[Time series graphs illustrating the multivariate ITY and MIT.]{Time series graphs (see definition in Sect.~\ref{sec:tsg}) illustrating the causal link-based multivariate ITY (a), and MIT (b). The respective CMIs are between $X_{t-\tau}$ and $Y_t$ (marked by the black dots) conditioned on the sets marked by the colored boxes.
% }
% %entropy $H(Y| \mathcal{P}_{Y} {\setminus}\{X\},\mathcal{P}_{X} )$ 
% \label{fig:entropy_flow}
% \end{figure}

\emph{Transfer entropy} (TE), introduced by Schreiber \cite{Schreiber2000b}, is the information-theoretic analogue of Granger causality and for multivariate Gaussian processes they can be shown to be equivalent \cite{Barnett2009}. The key idea to arrive at a causal notion of information transfer is to measure the information content of the past of a process $X$ at times $t'<t$ about the target variable $Y$ at time $t$ and exclude information from the common history shared by $X$ and $Y$. In its multivariate version, TE is defined as

\begin{align} \label{eq:def_te}
I^{\rm TE}_{X {\to} Y} &= I(X_t^-;Y_t\,|\,\mathbf{X}_t^- {\setminus} X_t^-)\,.
\end{align}

TE measures the aggregated influence of $X$ at all past lags, i.e., it is not lag-specific, and leads to the problem that infinite-dimensional densities have to be estimated, which is commonly called the ``curse of dimensionality''. In Ref.~\cite{Runge2012prl} this problem is overcome by a decomposition formula. In practice, however, a truncated version at some maximal delay is typically used. In Ref.~\cite{Runge2012b} a lag-specific variant of TE taking into account the time series graph structure was introduced, called the \emph{information transfer to Y} (ITY) defined as

\begin{align} \label{eq:def_ity}
I_{X \to Y}^{\rm ITY}(\tau) 
&= I(X_{t-\tau};Y_t|\mathcal{P}_{Y_t}{\setminus}\{X_{t-\tau}\})
% &= H(Y_t|\mathcal{P}_{Y_t}{\setminus}\{X_{t-\tau}\}) - H(Y_t|\mathcal{P}_{Y_t}),
\end{align}

ITY is different from a bivariate lag-specific TE definition such as in Ref.~\cite{Wibral2013a} since it explicitly uses the previously reconstructed parents $\mathcal{P}_{Y_t}\subset\mathbf{X}^-$, which includes drivers from the past of the whole process and not only $Y$'s own past.
 % In analogy, one can also define a contemporaneous ITY \cite{Runge2014d}.

TE can be derived as one component of decomposing the \emph{prediction entropy} $I(\mathbf{X}^-_t;Y_t)$ \cite{Faes2015a}. A similar approach is developed in Ref.~\cite{Stramaglia2012}.
The decisive difference of these transfer entropy related measures to our proposed framework is that they measure the contribution of different drivers to predicting a target variable $Y$, i.e., they are aimed at decomposing the \emph{predictive information}. In particular, Granger causality, TE or ITY are zero for indirect causal interactions, i.e., if the interaction is mediated via another measured process.
With respect to time series graphs, ITY is one way to quantify the strength of a causal coupling link between $X$ and $Y$ at some lag $\tau$. For a detailed account on the interpretability of different measures of the strength of causal links see Ref.~\cite{Runge2012b}.

\subsection{Quantifying information transfer along paths}
\label{sec:information_paths}
%%%%%%%%%%%%%%%%%%%%%%%%%%%%%%%%%%%%%%%%%%%%%%%%%%%%%%%%%%%%%%%%%%%%%%%%%%%%%%%%%%%%%%%%%%%%%%
\begin{figure*}[t!]
\begin{center}
\includegraphics[width=2\columnwidth]{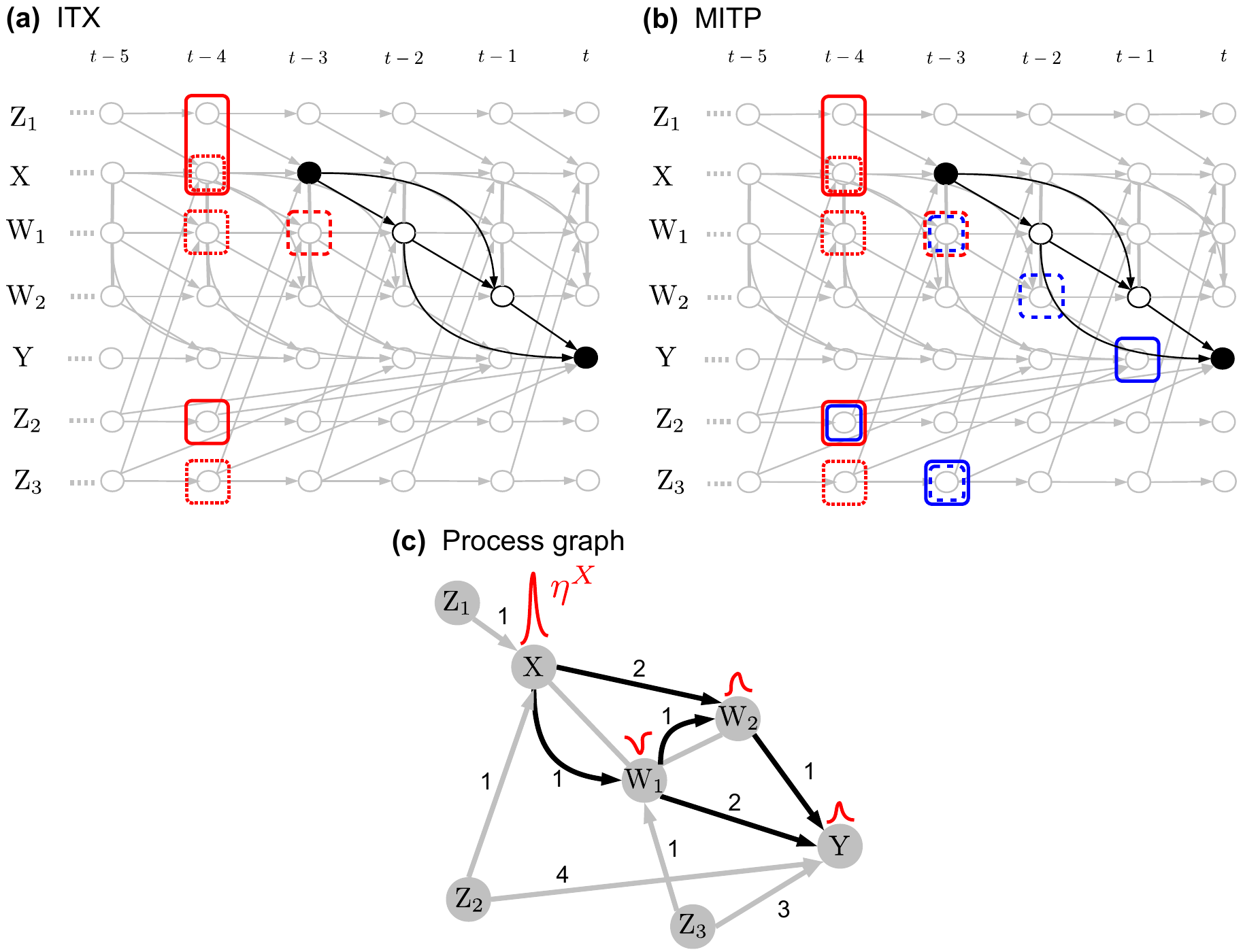}
\end{center}
\caption[Time series graph and process graph.]{(Color online) Time series graphs illustrating the path-based measures of information transfer ITX (a) and MITP (b), and process graph (c, the labels denote the lags). Directed links (Def.~\ref{eq:def_graph}) are marked by arrows, contemporaneous links (Def.~\ref{eq:def_graph_contemp}) by a solid line. There are three causal directed paths connecting $X_{t-3}$ and $Y_t$ (black lines), two of length $2$ via $W_{1,t-2}$ and $W_{2,t-1}$ and one of length $3$: $X_{t-3}\to W_{1,t-2}\to W_{2,t-1}\to Y_t$. The idea of the measure ITX shown in (a) here is to quantify how much of the information entering the system in $X_{t-3}$, i.e., the dynamical noise $\eta^X$, is transferred along causal paths to $Y_t$ by conditioning out the effect of the parents of $\mathcal{P}_{X_{t-3}}$ (solid red boxes), its neighbors involving contemporaneous sidepaths to $Y_t$ denoted $\mathcal{N}^{Y_t}_{X_{t-3}}$ (dashed red box), and the neighbor's parents $\mathcal{P}(\mathcal{N}^{Y_t}_{X_{t-3}})$ (dotted red boxes). The latter two conditioning sets exclude contemporaneous sidepaths like $X_{t-3}\,{\--}W_{1,t-3}\,\to W_{2,t-2}\,\to Y_{t-1}\,\to Y_{t}$. 
ITX still depends on processes affecting intermediate nodes on causal paths, e.g., process $Z_3$ which drives $W_1$ and $Y$. The idea of MITP shown in (b) now is to go one step further and isolate all causal paths from the remaining process by additionally conditioning on the parents of the intermediate path nodes $\mathcal{C}_{X_{t-\tau}{\to}Y_t}{\setminus}\{X_{t-\tau} \}$ (dashed blue boxes) and $Y$ (solid blue boxes). This also allows to isolate mediated effects using momentary interaction information as defined in Sect.~\ref{sec:def_interactions}.
}
%entropy $H(Y| \mathcal{P}_{Y} {\setminus}\{X\},\mathcal{P}_{X} )$ 
\label{fig:path_measures}
\end{figure*}

In this article the main question of interest is not only how strong a causal link is, but more generally how strong an indirect causal influence of a variable $X_{t-\tau}$ on $Y_t$ is (Fig.~\ref{fig:path_measures}). Indirect causal effects can only be transferred on \emph{causal paths} in the time series graph, which are paths consisting only of directed links as defined in Sect.~\ref{sec:tsg_paths}.
Note that Fig.~\ref{fig:path_measures}(c) shows an aggregated process graph, which is not suited to read off causal paths since it does not show the full spatio-temporal causal structure (including autodependencies) like time series graphs.

We denote the processes along causal paths including $X_{t-\tau}$ for $\tau>0$ and excluding $Y_t$ by

\begin{align} \label{eq:causal_paths}
\mathcal{C}_{X_{t-\tau}\to Y_t} =& \{X_{t-\tau}\} \cup\nonumber\\
&  \{W_{t-\tau_W} \in \mathbf{X}^-_t \text{ with $\tau>\tau_W>0$}:\nonumber\\
&X_{t-\tau}\to \ldots \to W_{t-\tau_W} \to \ldots \to Y_t\}\,, 
\end{align}

where $\to \ldots \to$ denotes a succession of directed links or only one directed link.
These can be read off directly from the time series graph. For example, in Fig.~\ref{fig:path_measures}, $X_{t-3}$ and $Y_t$ are connected by the three causal paths $X_{t-3}\to W_{2,t-1}\to  Y_{t}$, $X_{t-3}\to W_{1,t-2}\to  Y_{t}$, and $X_{t-3}\to W_{1,t-2}\to W_{2,t-1}\to  Y_{t}$ such that $\mathcal{C}_{X_{t-3}\to Y_t}=\{ X_{t-3},\,W_{1,t-2},\,W_{2,t-1} \}$. 
Our goal is now to construct a CMI with conditions that leave open only these causal paths and block all non-causal paths according to the definition of paths and blocking in time series graphs in Sect.~\ref{sec:tsg_paths}. 

The first step is to exclude paths due to common drivers of $X$ and $Y$. The parents $\mathcal{P}_{X_{t-\tau}}$ of $X$ at time $t-\tau$ block all common drivers from the past since these paths necessarily contain the motifs ${\--}\,\bullet\,\to X_{t-\tau}$ or $\to\,\bullet\,\to X_{t-\tau}$, which are both blocked if conditioned on. 
A second class of non-causal paths are contemporaneous sidepaths as defined in Sect.~\ref{sec:tsg_paths}. These can be blocked by conditioning on those contemporaneous neighbors of $X_{t-\tau}$ that have at least one contemporaneous sidepath, of course \emph{not traversing} $X_{t-\tau}$, which we define as

\begin{align} \label{eq:sidepath_neighbors}
\mathcal{N}^{Y_t}_{X_{t-\tau}} = \{W_{t-\tau} \in \mathcal{N}_{X_{t-\tau}}:~X_{t-\tau}\,{\--}\,W_{t-\tau}\, \substack{\to\\ {\--}}\, \ldots\to Y_t\}\,,
\end{align}

where $ \substack{\to\\ {\--}}\, \ldots \to$ denotes either a directed path or a contemporaneous sidepath that does not involve $X_{t-\tau}$.
For example, in Fig.~\ref{fig:path_measures}(a,b), $\mathcal{N}^{Y_t}_{X_{t-3}}=\{ W_{1,t-3}\}$. On the other hand, for the causal path $X_{t-2}\to X_{t-1} \to X_t$ we have $\mathcal{N}^{X_t}_{X_{t-2}}=\emptyset$, since there are no contemporaneous sidepaths from $ W_{1,t-2}$ to $X_t$.
The condition on neighbors unfortunately introduces new open paths because $X_{t-\tau}\,{\--}\,\blacksquare \leftarrow$ is an open motif. To block these paths, one needs to additionally condition on the parents of the neighbors $\mathcal{P}(\mathcal{N}^{Y_t}_{X_{t-\tau}})$.
Note that one could also only select those parents from $X_{t-\tau}$ which have a `common driver path' to $Y_t$, but our goal is to isolate the momentary information \emph{entering the system in $X$}, i.e., the dynamical noise from model~(\ref{eq:source_process}), and quantify its propagation along causal paths to $Y$ some time later. 
The \emph{information transfer from $X$} (ITX) is now defined for $\tau>0$ as

\begin{align} \label{eq:def_itx}
I^{\rm ITX}_{X_{t-\tau}\to Y_t} = I(X_{t-\tau};Y_t|\mathcal{P}_{X_{t-\tau}},\,\mathcal{N}^{Y_t}_{X_{t-\tau}},\,\mathcal{P}(\mathcal{N}^{Y_t}_{X_{t-\tau}}))\,.
\end{align}

It measures the part of source entropy in $X_{t{-}\tau}$ that reaches $Y_t$ on any causal path and could be regarded as an information-theoretic analogue to Sims causality as mentioned in the introduction (see also Table~\ref{tab:tigramite}). In Ref.~\cite{Runge2012b} this measure was introduced without the condition on neighbors.

ITX does not exclude information entering process $Y_t$ from other sources, for example from process $Z_3$ in the example shown in Fig.~\ref{fig:path_measures}(a). The idea of momentary information transfer \cite{Runge2012b} was to isolate the information shared between two processes via a causal link from the remaining process. Now this idea can be generalized by isolating all causal paths from the remaining process to assess the part of the source entropy of $X_{t-\tau}$ that is transferred on any causal path and shared with $Y_t$, excluding the parents of all intermediate path nodes and $Y$ that are not part of the causal path. Figure~\ref{fig:path_measures}(b) illustrates this idea. With the nodes on all causal paths including $X_{t-\tau}$  denoted by $\mathcal{C}_{X_{t-\tau}\to Y_t}$ (Eq.~(\ref{eq:causal_paths})), the \emph{momentary information transfer along causal paths} (MITP) is defined as

\begin{align} \label{eq:def_mitp}
I^{\rm MITP}_{X{\to}Y}(\tau)&= I(X_{t-\tau};Y_t~|~ \mathcal{P}_{Y_t}{\setminus}\mathcal{C}_{X_{t-\tau}{\to}Y_t},  \mathcal{P}(\mathcal{C}_{X_{t-\tau}{\to}Y_t}),\nonumber\\
&\phantom{= I(X_{t-\tau};Y_t~|~}\mathcal{N}^{Y_t}_{X_{t-\tau}},\,\mathcal{P}(\mathcal{N}^{Y_t}_{X_{t-\tau}}) ).
\end{align}

For the time series graph example in Fig.~\ref{fig:path_measures}(b), these conditions are marked by the red and blue boxes. In Sect.~\ref{sec:analytical_examples} we will prove that MITP, contrary to ITX, also fulfills a generalized coupling strength autonomy theorem which allows to better relate it to the underlying dynamics of a process as will be discussed in Sect.~\ref{sec:discussion}.

If $\mathcal{C}_{X_{t-\tau}\to Y_t}=\{X_{t-\tau}\}$, and under the ``no sidepath''-constraint in Ref.~\cite{Runge2012b}, the conditions on the neighbors can be dropped and MITP collapses to MIT.

\subsection{Quantifying mediating information transfer}
\label{sec:def_interactions}

Looking at Fig.~\ref{fig:path_measures}, one immediate question is whether one can quantify how much of the information transfer between $X$ and $Y$ went through $W_1$ and how much through $W_2$? Which of these is information-theoretically more important for explaining the indirect causal relationship between $X$ and $Y$? The interaction information defined in Eq.~(\ref{eq:def_interaction_information}) can be used to answer this question, we here discuss two analogous versions for the measures ITX and MITP. For two processes $X_{t-\tau}$ and $Y_t$ connected by a causal path, intermediate processes can occur with multiple lags. For example, among the causal paths between $X_{t-4}$ and $Y_t$ in Fig.~\ref{fig:path_measures}, the process $W_1$ is traversed at lags $W_{1,t-2}$ and $W_{1,t-3}$. Generally, if a subprocess $W$ is intermediate in an interaction $X_{t-\tau}\to\cdots \to Y_t$ at multiple lags $t-\tau_1,\,t-\tau_2,\ldots$, we here include all these lags in the vector $\mathbf{W}=\{W_{t-\tau_1},W_{t-\tau_2},\ldots\}\subset \mathcal{C}_{X_{t-\tau}\to Y_t}$. 

First, we define the \emph{interaction information from X} (IIX) as

\begin{align} \label{eq:interaction-itx}
&\mathcal{I}^{\rm IIX}_{X_{t-\tau}\to Y_t|\mathbf{W}} \nonumber\\
&= \mathcal{I}(X_{t-\tau};Y_t;\mathbf{W}~|~ \mathcal{P}_{X_{t-\tau}},\,\mathcal{N}^{Y_t}_{X_{t-\tau}},\,\mathcal{P}(\mathcal{N}^{Y_t}_{X_{t-\tau}}) ) \\
%&= \mathcal{I}(X_{t-\tau};Y_t;W_{t-\tau_W}|\mathcal{P}_{X_{t-\tau}}) \nonumber\\
 &= I^{\rm ITX}_{X{\to}Y}(\tau) - \underbrace{I(X_{t-\tau};Y_t|\mathcal{P}_{X_{t-\tau}},\mathcal{N}^{Y_t}_{X_{t-\tau}},\,\mathcal{P}(\mathcal{N}^{Y_t}_{X_{t-\tau}}),\,\mathbf{W} )}_{\text{ITX conditioned on $\mathbf{W}$}}\,.
\end{align}

IIX measures the effect of an intermediate process $\mathbf{W}$ on the information transfer between the source information of $X_{t-\tau}$ and $Y_t$. 
Second, the \emph{momentary interaction information} (MII) for an intermediate process $\mathbf{W}$ is defined as

\begin{align} \label{eq:def_mii}
&\mathcal{I}^{\rm MII}_{X{\to}Y|\mathbf{W}}(\tau) \nonumber\\
&= \mathcal{I}(X_{t-\tau};Y_t;\mathbf{W}~|~\mathcal{P}_{Y_t}{\setminus}\mathcal{C}_{X_{t-\tau}{\to}Y_t},  \mathcal{P}(\mathcal{C}_{X_{t-\tau}{\to}Y_t}),\nonumber\\
&\phantom{= \mathcal{I}(X_{t-\tau};Y_t;\mathbf{W}~|~}\mathcal{N}^{Y_t}_{X_{t-\tau}},\,\mathcal{P}(\mathcal{N}^{Y_t}_{X_{t-\tau}}) ) \\
&=I^{\rm MITP}_{X{\to}Y}(\tau) - I(X_{t-\tau};Y_t~|~  \mathcal{P}_{Y_t}{\setminus}\mathcal{C}_{X_{t-\tau}{\to}Y_t},  \mathcal{P}(\mathcal{C}_{X_{t-\tau}{\to}Y_t}),\nonumber\\
&\phantom{I^{\rm MITP}_{X{\to}Y}(\tau)-} \underbrace{\phantom{I(X_{t-\tau};Y_t~|~}\mathcal{N}^{Y_t}_{X_{t-\tau}},\,\mathcal{P}(\mathcal{N}^{Y_t}_{X_{t-\tau}}), \mathbf{W} )~~~~~~~~~~}_{\text{MITP conditioned on $\mathbf{W}$}}\,.
\end{align}

MII measures the effect of $\mathbf{W}$ on the momentary information transfer along paths between $X_{t-\tau}$ and $Y_t$ and additionally isolates the influence of drivers of the causal path processes.
In Section~\ref{sec:analytical_examples} we discuss several examples demonstrating that IIX and MII are not necessarily always positive implying that an intermediate process can counteract the interaction between $X_{t-\tau}$ and $Y_t$. This measure can naturally be extended by including sets of processes from $\mathcal{C}_{X_{t-\tau}\to Y_t}$.
Due to the symmetry of interaction information as defined in Eq.~(\ref{eq:def_interaction_information}),  MII is symmetric in its arguments excluding the condition. 

Table~\ref{tab:tigramite} provides an overview over the different classes of measures discussed here.
In a climate data example in Sect.~\ref{sec:climate}, we will see how IIX and MII can be used to quantify dominant pathway mechanisms and in Sect.~\ref{sec:information_efficiency} we discuss how they can be used as an aggregate measure of `causal interaction betweenness', modifying concepts from complex network theory for functional network analysis \cite{Bullmore2009}.

%%%%%%%%%%%%%%%%%%%%%%%%%%%%%%%%%%%%%%%%%%%%%%%%%%%%%%%%%%%%%%%%%%%%
\section{Examples and theorems} \label{sec:analytical_examples}

In the following we discuss how the novel approach allows to extract a detailed picture of interaction mechanisms between multiple processes. 
% In particular, we show that the goal of our approach is not a decomposition of predictive information about a target variable $Y$ as in previous works \cite{Faes2015,Stramaglia2012,Stramaglia2015}, but a quantification of different parts of an interaction mechanism between two processes $X$ and $Y$.

\subsection{Linear model example}
\label{sec:sidepaths}
%%%%%%%%%%%%%%%%%%%%%%%%%%%%%%%%%
\begin{figure*}[!t]
\begin{center}
\includegraphics[width=1.5\columnwidth]{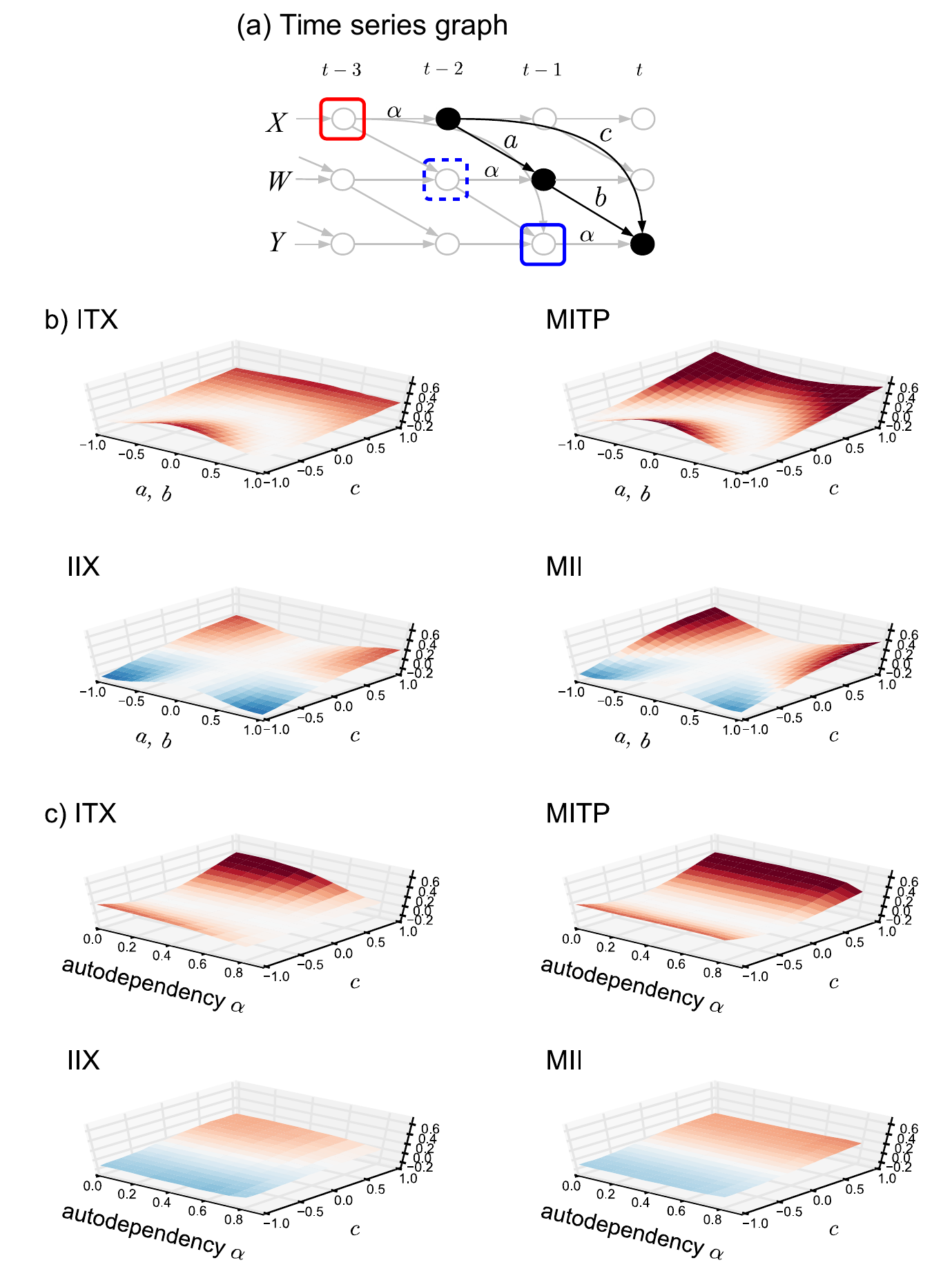}
\end{center}
\caption[Example time series graphs for illustrating momentary interaction information.]{(Color online) (a) Time series graph for model~(\ref{eq:xwy_model}) of a causal interaction between three processes at different lags (black dots). The parents are shown in colored boxes, here there are no neighbors. $a,\,b,\,c,\,\alpha$ denote the model coefficients. In (b) the interaction measures are plotted against $a=b$ (strength of the sidepath) and $c$ (strength of direct link) for an autodependency strength $\alpha=0.5$, and in (c) against $c$ and the autodependency strength $\alpha$ for  $a=b=0.5$. The color shading only emphasizes the sign and strength, the value can be read off the $z$-axis. All innovation terms $\eta$ have unit variance. Further parameters: ensemble size $30$, sample length $T=10,000$,  nearest-neighbor estimation parameter $k=1$.
}
\label{fig:interaction_motifs}
\end{figure*}
%%%%%%%%%%%%%%%%%%%%%%%%%%%%%%%%%%%%

In Ref.~\cite{Runge2012b} the strength of direct causal links was studied. The main finding was that MIT solely depends on the coefficient corresponding to the causal link. This property was called \emph{coupling strength autonomy} in Ref.~\cite{Runge2012b} and will be reviewed in Sect.~\ref{sec:coupling_strength_autonomy_general}.
For the case of interactions along causal paths, consider the following linear model with time series graph visualized in Fig.~\ref{fig:interaction_motifs}(a):

\begin{align} \label{eq:xwy_model}
X_t &= \alpha X_{t-1} + \eta^X_{t} \nonumber  \\
W_t &= \alpha W_{t-1} + a X_{t-1} + \eta^W_t  \nonumber  \\
Y_t &= \alpha Y_{t-1} + c X_{t-2} + b W_{t-1} + \eta^Y_t \,,
\end{align}

where all processes are jointly zero-mean Gaussian with variances $\sigma_X^2,\,\sigma_Y^2,\,\sigma_Z^2$ of the innovation terms $\eta^{\cdot}$. Here the influence of $X_{t-2}$ on $Y_t$ has two paths: One via the direct coupling link $X_{t-2}\to~Y_t$ and one via the path $X_{t-2}\to W_{t-1}\to Y_t$ such that we can rewrite 

\begin{align}
Y_t &= c X_{t-2} + b (a X_{t-2} + \eta^W_{t-1}) + \eta^Y_t,
\end{align}

from which we see that the coupling cannot be unambiguously related to one coefficient and interesting dynamics emerge. 
In Fig.~\ref{fig:interaction_motifs}(b) we investigate the measures ITX, MITP, IIX, and MII numerically for varying $a=b$ (strength of the sidepath) and $c$ (strength of direct link) for fixed autodependency strength $\alpha=0.5$. We assume $a,b\neq 0$, because otherwise this causal path vanishes and IIX or MII are not defined. The ensemble size to estimate the ensemble mean is $30$, the sample length is $T=10,000$, and the CMI nearest-neighbor estimation parameter is $k=1$ to achieve minimal bias \cite{Frenzel2007}. As mentioned in Sect.~\ref{sec:cmi_estimation}, for larger $k$ the bias increases, but also the estimator's variance decreases \cite{Frenzel2007} making higher $k$ values a better choice for independence tests as used in the causal algorithm \cite{Runge2012prl}.

Since we vary $a$ together with $b$, the contribution via this sidepath is always positive, also for negative $a,\,b$. If also $c$ is positive, we observe an increase in ITX as well as MITP (Fig.~\ref{fig:interaction_motifs}(b)), with the latter being more pronounced. For negative $c$, on the other hand, the contributions of the direct link and the sidepath \emph{counteract} and, for certain values $(a,\,b,\,c)$ even cancel out leading to a vanishing ITX and MITP.

These different types of mediation of the intermediate process $W$ can be quantified by IIX and MII (lower panels in Fig.~\ref{fig:interaction_motifs}(b)): For positive $c$, both are larger than zero, showing the positive contribution of both mechanisms, also here MII is more pronounced. For $c=0$, MII is equal to MITP because the only interaction stems from the causal path demonstrating the explanatory influence of $W$, which acts as the only mediating process. In the Venn diagram of Fig.~\ref{fig:info_theory}(c) this corresponds to the case in which $H(W)$ entails all of the shared entropy between $X$ and $Y$.
For negative $c$, the counteracting effect is evident in the negative sign of IIX and MII which implies for the latter that $\mathcal{I}^{\rm MII}_{X\to Y|W}(\tau=2)>I^{\rm MITP}_{X\to Y}(\tau=2)$: Conditioning out the effect of the intermediate process $W$ here reveals that the direct link is actually very strong and was only `masked' by the counteracting sidepath via $W$.  In Ref.~\cite{Stramaglia2012} a similar case, but without isolating the interaction pathway, was termed a ``synergistic'' contribution to the predictive information about $Y$ as opposed to the ``redundant'' case with a positive interaction information.

In Fig.~\ref{fig:interaction_motifs}(c) the dependence of the four measures for $a=b=0.5$ and varying the autodependency strength $\alpha$ and direct link strength $c$ is shown. ITX features a strong dependency on $\alpha$ already for weak drivings $\alpha \approx 0.4$ and almost vanishes for a very strong driving. Note that the same effect would be observed if other external processes drive $W$ and $Y$ (from $X$ the effect is partially excluded due to the condition on $\mathcal{P}_X$). 
Analytically, here ITX can only be reduced to 

\begin{align} \label{eq:ana_itx}
I^{\rm ITX}_{X\to Y} (\tau=2) &\overset{\phantom{\text{Eq.~(\ref{eq:cmi_trans_inv}) in Appendix}}}{=} I\left(X_{t-2};Y_t|X_{t-3}\right) \nonumber\\
&\overset{\phantom{\text{Eq.~(\ref{eq:cmi_trans_inv}) in Appendix}}}{=} I\left(\alpha X_{t-3} + \eta^X_{t-2};Y_t|X_{t-3}\right) \nonumber\\
                            &\overset{\text{Eq.~(\ref{eq:cmi_trans_inv}) in Appendix}}{=} I\left(\eta^X_{t-2};Y_t|X_{t-3}\right),
\end{align}

which still depends on many coefficients in the model and cannot be easily related to the underlying dynamics. On the other hand, MITP can be simply related to the coefficients along the causal paths as

\begin{align} \label{eq:ana_mitp_sidepath}
I^{\rm MITP}_{X\to Y}(\tau=2) &=\frac{1}{2} \ln \left(1+\frac{(c+a b)^2 \sigma_X^2}{b^2 \sigma_W^2+\sigma_Y^2}\right)\,,
\end{align}

which follows from Theorem~\ref{thm:autonomy_mitp} in Sect.~\ref{sec:coupling_strength_autonomy_general}. Here it becomes evident that MITP vanishes along the parabola $c=-a\,b$ (which can be considered a pathological case where the causal assumption of \emph{faithfulness} is violated \cite{Spirtes2000}). A second important finding is that MITP is \emph{independent} of the autodependency coefficient $\alpha$. The same holds for MII, here given by

\begin{align} \label{eq:interaction_example_chain}
\mathcal{I}^{\rm MII}_{X{\to}Y|W}  =& \frac{1}{2} \ln \left(1+\frac{(c+a b)^2 \sigma_X^2}{b^2 \sigma_W^2+\sigma_Y^2}\right) \nonumber\\
&- \frac{1}{2} \ln \left(1+\frac{c^2 \sigma_X^2 \sigma_W^2}{(\sigma_W^2+a^2\sigma_X^2)\sigma_Y^2}\right)\,,
\end{align}

as follows from Appendix~\ref{sec:app_theorem_mii}. 
This implies that the value of MITP and MII can solely be related to the model's coefficients along the causal interaction paths, which can be considered an advantage in interpreting these measures compared to ITX or IIX. While in this example there are no external parents influencing the processes along the path, in more complex schemes also their effect can be excluded by the condition on the parents of the nodes on the path denoted by $\mathcal{C}_{X_{t-\tau}\to Y_t}$. In Sect.~\ref{sec:coupling_strength_autonomy_mii} this will be proven for the general case.
Note that in Fig.~\ref{fig:interaction_motifs}(c) MITP and MII are slightly affected for very strong autodependencies which is due to an estimation bias and vanishes for infinite sample sizes.
This model will be further discussed in relation to linear causal effect measures in Sect.~\ref{sec:causal_effect}.
% In Sect.~\ref{sec:climate} we discuss a climatological application.

\subsection{Nonlinear model example}
\label{sec:interaction_examples}
%%%%%%%%%%%%%%%%%%%%%%%%%%%%%%%%%
\begin{figure*}[!t]
\begin{center}
\includegraphics[width=1.5\columnwidth]{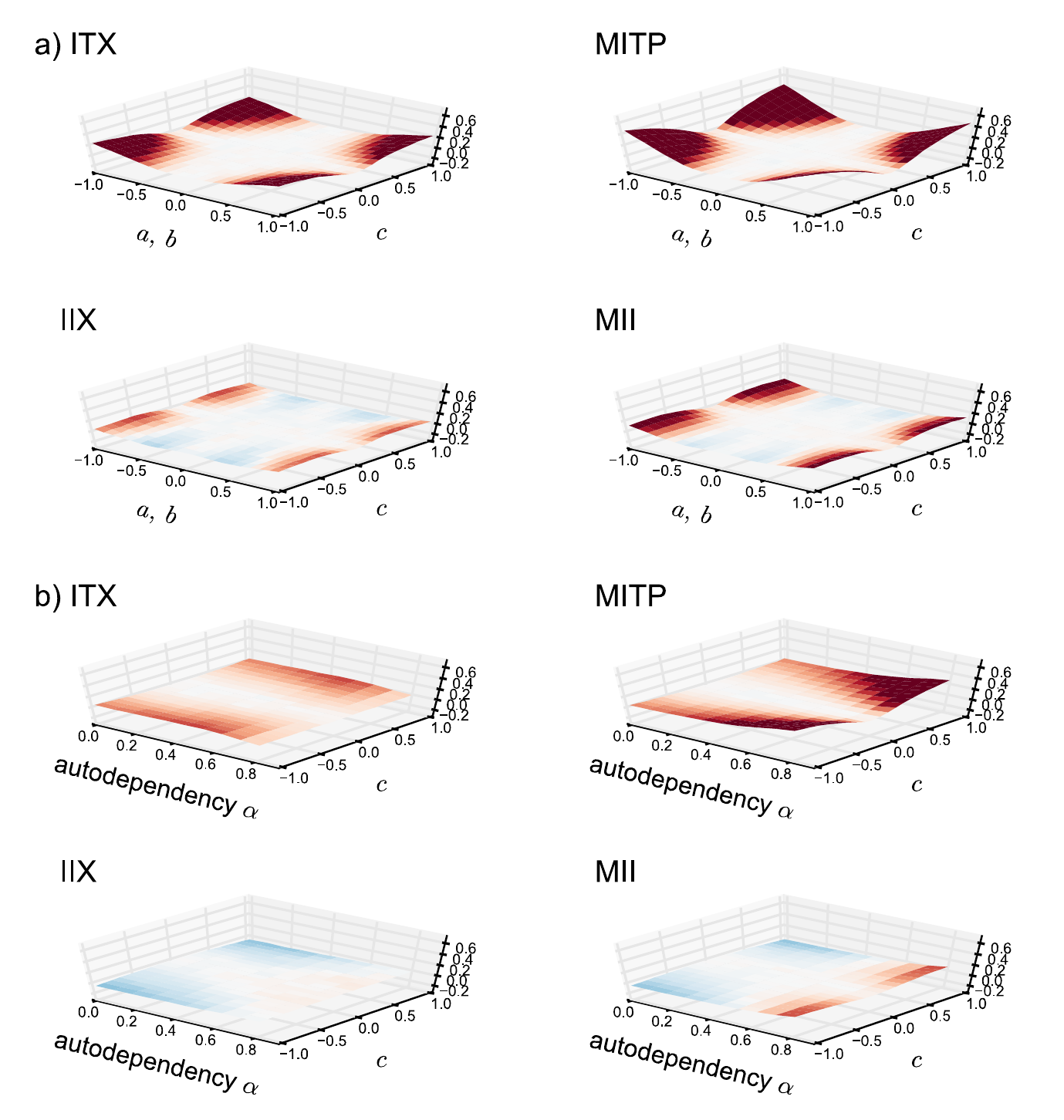}
\end{center}
\caption[Example time series graphs for illustrating momentary interaction information.]{(Color online) Same as in Fig.~\ref{fig:interaction_motifs}, but for the nonlinear model~(\ref{eq:xwy_model_synergetic}).}
\label{fig:interaction_motifs_synergetic}
\end{figure*}
%%%%%%%%%%%%%%%%%%%%%%%%%%%%%%%%%%%%

Next, we discuss a nonlinear version of model~(\ref{eq:xwy_model}) which shares the same time series graph, but features different dynamics:

\begin{align} \label{eq:xwy_model_synergetic}
X_t &= \alpha X_{t-1} + \eta^X_{t} \nonumber  \\
W_t &= \alpha W_{t-1} + a X_{t-1} + \eta^W_t  \nonumber  \\
Y_t &= \alpha Y_{t-1} + \underbrace{c~b~X_{t-2}~W_{t-1}}_{\text{multiplicative dependency}} +  \eta^Y_t \,,
\end{align}

with Gaussian innovation terms as before. Figure~\ref{fig:interaction_motifs_synergetic}(a) shows that ITX and MITP vanish for $b$ or $c$ equal zero and are increasing for larger absolute values. For larger $|c|$ and certain values of $a,b$ we observe a counteracting of $W$ through the indirect path as can be seen from the negative IIX and MII, but no annihilation of both effects occurs here and ITX and MITP stay positive.
For this nonlinear dependency structure both ITX and MITP (and the corresponding interaction informations) depend on the external forcing parameter $\alpha$ (Fig.~\ref{fig:interaction_motifs_synergetic}(b)). The reason is that the nonlinearity mixes the terms and the dependencies cannot be conditioned out anymore. Consider model~(\ref{eq:xwy_model_synergetic}), but with differing autodependency terms $\alpha,\,\beta,\,\gamma$ for $X,\,W,\,Y$, respectively.
MITP here is given by

\begin{align} 
I^{\rm MITP}_{X\to Y} (\tau=2) &\overset{\phantom{\text{Eq.~(\ref{eq:cmi_trans_inv})}}}{=} I\left(X_{t-2};Y_t|X_{t-3},\,W_{t-2},\,Y_{t-1}\right) \nonumber\\
                            &\overset{\text{Eq.~(\ref{eq:cmi_trans_inv})}}{=} I\left(\eta^X_{t-2};Y_t|X_{t-3},\,W_{t-2},\,Y_{t-1}\right)\,,
\end{align}

and the dependency of $Y_t$ can be rewritten as

\begin{align}
Y_t =& c b (a \eta^X_{t-2} +  \eta^W_{t-1}) \eta^X_{t-2} +  \eta^Y_t \nonumber \\
    & + cb (\alpha a X_{t-3} \eta^X_{t-2} + \alpha X_{t-3} \eta^W_{t-1} \nonumber\\
    &\phantom{+ cb (}+ \beta W_{t-2} \eta^X_{t-2} + \alpha a X_{t-3} \eta^X_{t-2}) \nonumber\\
    & + \gamma Y_{t-1} + c b (\alpha \beta X_{t-3} W_{t-2} + a \alpha^2 X_{t-3}^2)\,.
\end{align}

Here in MITP the last line vanishes due to the condition on $(X_{t-3},\,W_{t-2},\,Y_{t-1})$, but due to the multiplicative mixing with the noise terms in the second and third line, the autodependency coefficients $\alpha,\,\beta$ (but not $\gamma$) still determine MITP. ITX additionally depends on $\gamma$.
This model, therefore, demonstrates a case where `external effects' cannot be excluded anymore. Thus, while the information-theoretic interpretation still holds, MITP cannot be easily related to the system's dynamics.
Still, plots like in Figs.~\ref{fig:interaction_motifs},\,\ref{fig:interaction_motifs_synergetic} can help to better understand dynamical interactions also in toy models from nonlinear dynamics. In the next section we prove under which general assumptions the coupling strength autonomy holds for MITP and MII.
The multiplicative dependency can be seen as an example of synergy which has recently gained a lot of interest in information-theoretical studies, see e.g. Refs.~\cite{Olbrich2015,Barrett2015}. In Ref.~\cite{Runge2014b} synergistic effects are studied with respect to optimal prediction schemes.

\subsection{Theorems}
\label{sec:coupling_strength_autonomy_general}

In this section we state some inequality relations among the novel measures and generalize the coupling strength autonomy theorem for MIT \cite{Runge2012b} to the path-based measures MITP and MII.

\begin{Theorem}[Inequality relations] \label{thm:inequality_mitp}
For $\tau>0$, the following inequalities hold:

\begin{align}
 % I_{X \to Y}^{\rm ITX}(\tau) &\leq I^{\rm MIT}_{X{\to}Y}(\tau) \leq  I_{X \to Y}^{\rm ITY}(\tau)\\
\mathcal{I}_{X \to Y|\mathbf{W}}^{\rm IIX}(\tau) &\leq I^{\rm ITX}_{X{\to}Y}(\tau) \\
\mathcal{I}_{X \to Y|\mathbf{W}}^{\rm MII}(\tau) &\leq I^{\rm MITP}_{X{\to}Y}(\tau) \\
 I_{X \to Y}^{\rm ITX}(\tau) &\leq I^{\rm MITP}_{X{\to}Y}(\tau)\,.
\end{align}

\end{Theorem}

The first two inequalities are trivially fulfilled since IIX and MII are defined by ITX and MITP \emph{minus} a CMI, which is always positive. Equality holds if the intermediate node(s) $\mathbf{W}$ explain the entire interaction between $X$ and $Y$. The last inequality is proven in appendix~\ref{thm:inequality_mitp_proof}.
 % holds only under the ``no contemporaneous dependency''-condition~(\ref{cond:nosidepath_contemp}). 
In practice, this inequality is often not fulfilled because the estimation dimension of MITP is typically much larger than that of ITX and finite sample effects lead to a negative bias which often leads to MITP being smaller than ITX. This also makes a comparison of the values of ITX and MITP more difficult.

% \subsubsection{Momentary information transfer along paths (MITP)}
To generalize the coupling strength autonomy theorem from MIT to MITP and MII, we consider causal paths as defined in Sect.~\ref{sec:tsg_paths} instead of only causal links.
While the careful condition on only those neighbors that have sidepaths excludes dependencies of MITP and MII on the dynamics along these sidepaths, one cannot avoid a contemporaneous dependency on the interaction with the respective neighbor itself. This also holds for other intermediate processes on causal paths. 
For the following theorems, we define as a ``no contemporaneous dependency''-condition
% . Denoting by $\mathcal{S}^{\rm MITP}=(~\mathcal{P}_{Y_t}{\setminus}\mathcal{C}_{X_{t-\tau}{\to}Y_t},  \mathcal{P}(\mathcal{C}_{X_{t-\tau}{\to}Y_t}), \mathcal{N}^{Y_t}_{X_{t-\tau}},\,\mathcal{P}(\mathcal{N}^{Y_t}_{X_{t-\tau}}))$ the conditions in MITP, it can be expressed as
\begin{align} \label{cond:nosidepath_contemp}
 \forall~~ W^{(i)}_{t-\tau_i} \in \mathcal{C}_{X_{t-\tau}\to Y_t}: ~~ \mathcal{N}_{W^{(i)}_{t-\tau_i}}^{Y_t} = \emptyset
% &  I(W^{(i)}_{t-\tau_i}; \mathcal{S}^{\rm MITP}\setminus \mathcal{P}_{W^{(i)}_{t-\tau_i}}~|~\mathcal{P}_{W^{(i)}_{t-\tau_i}})=0\,,  
\end{align}
with $\mathcal{N}_{W^{(i)}_{t-\tau_i}}^{Y_t}$ defined in Eq.~(\ref{eq:sidepath_neighbors}). This condition implies that no contemporaneous sidepaths as defined in Sect.~\ref{sec:tsg_paths} emanate from any of the path nodes $\mathcal{C}_{X_{t-\tau}\to Y_t}$ (including $X_{t-\tau}$) towards $Y_t$.
% which essentially isolates the `dynamical noise' in each $W^{(i)}$ and tests for its contemporaneous independence with any of the conditions on external processes involved in MITP. 
Note that we denote by $W^{(i)}_{t-\tau_i}$ each individual subprocess along causal paths at a certain lag $\tau_i$. If one subprocess occurs at multiple lags, it will have another index $i$ for each lag.

\begin{Theorem}[Coupling strength autonomy for MITP] \label{thm:autonomy_mitp}
Let $X,\,Y$ be two subprocesses of a multivariate stationary discrete-time process $\mathbf{X}$ sufficing the Markov property (Eq.~(\ref{eq:markovity})) with time series graph $\mathcal{G}$. 
We assume that $X_{t-\tau}$ and $Y_t$ are connected by a directed path with path nodes $\mathcal{C}_{X_{t-\tau}\to Y_t}$ including $X_{t-\tau}$ as defined in Eq.~(\ref{eq:causal_paths}). 
We denote those parents of $Y_t$ that are in the path nodes as $\mathcal{P}^\mathcal{C}_Y=\mathcal{P}_{Y_t} \cap \mathcal{C}_{X_{t-\tau}\to Y_t}$ and correspondingly for other path nodes and assume the following dependencies:

\begin{align} \label{cond:additivity_paths}
 X_t &=  g_X(\mathcal{P}_{X_{t-\tau}}) + \eta^X_t \nonumber \\
 Y_t &=  f_Y(\mathcal{P}^\mathcal{C}_Y) + g_Y (\mathcal{P}_{Y_t}\setminus \mathcal{P}^\mathcal{C}_Y) +  \eta^Y_t,
\end{align}

where $f_Y$ is linear and $g_{X,Y}$ arbitrary. Further, for all path nodes $W^{(i)}$ we assume the dependencies

\begin{align} \label{cond:linearity_path_nodes}
W^{(i)}_t =&  f_i(\mathcal{P}^\mathcal{C}_i) + g_i (\mathcal{P}_i\setminus \mathcal{P}^\mathcal{C}_i) +  \eta^{i}_t \nonumber\\
&\forall~~ W^{(i)}\in\mathcal{C}_{X_{t-\tau}\to Y_t}\setminus\{X_{t-\tau}\},
\end{align}

where the $f_i$ are again linear, the $g_i$ are arbitrary functions and the dynamical noise terms $\eta^{\cdot}$ are i.i.d. due to Markovity. Then, MITP (Eq.~(\ref{eq:def_mitp})) is given by

\begin{align} \label{eq:mitp_theorem_sidepaths}
&I^{\rm MITP}_{X{\to}Y}(\tau) =\nonumber\\
&I(\eta^X_{t-\tau}~;~\eta^Y_t + f(\eta^X_{t-\tau}, \cup_{i} \eta^i_{t-\tau_i})\nonumber\\
&\phantom{  I(}|~\mathcal{P}_{Y_t}{\setminus}\mathcal{C}_{X_{t-\tau}{\to}Y_t},  \mathcal{P}(\mathcal{C}_{X_{t-\tau}{\to}Y_t}),\mathcal{N}^{Y_t}_{X_{t-\tau}},\,\mathcal{P}(\mathcal{N}^{Y_t}_{X_{t-\tau}}))\,,
\end{align}

where $0<\tau_i<\tau ~\forall~ i$.
If further the ``no contemporaneous dependency''-condition~(\ref{cond:nosidepath_contemp}) holds, MITP reduces to a mutual information

\begin{align} \label{eq:mitp_theorem}
I^{\rm MITP}_{X{\to}Y}(\tau) &=  I(\eta^X_{t-\tau}~;~\eta^Y_t + f(\eta^X_{t-\tau}, \cup_{i} \eta^i_{t-\tau_i})),
\end{align}

where $f$ is a linear function and $\cup_{i} \eta^i$ denotes the innovation terms or dynamical noise of all path nodes in $\mathcal{C}_{X_{t-\tau}\to Y_t}\setminus\{X_{t-\tau}\}$.
\end{Theorem}

The proof is given in Appendix~\ref{sec:app_theorem_mitp}. This theorem also includes the coupling strength autonomy theorem for MIT \cite{Runge2012b} as a special case if $\mathcal{C}_{X_{t-\tau}\to Y_t}=\{X_{t-\tau}\}$ and under the ``no sidepath''-constraint in Ref.~\cite{Runge2012b}, then $f(\eta^X_{t-\tau}, \cup_{i} \eta^i_{t-\tau_i})=f(\eta^X_{t-\tau})$.

Since momentary interaction information (MII) is the difference between MITP and the MITP conditioned on one of the path nodes (excluding $X_{t-\tau}$), the theorem follows from the above theorem. 

\begin{Theorem}[Coupling strength autonomy for MII] \label{sec:coupling_strength_autonomy_mii}
Using the same assumptions as for Theorem~\ref{thm:autonomy_mitp},  the momentary interaction information $\mathcal{I}^{\rm MII}_{X{\to}Y|\mathbf{W}}(\tau)$ between $X_{t-\tau}$, $Y_t$ and one or more intermediate processes $\mathbf{W}=(W^{(1)}_{t-\tau_{1}},\,W^{(2)}_{t-\tau_{2}}\,\ldots)\in \mathcal{C}_{X_{t-\tau}\to Y_t}\setminus\{X_{t-\tau}\}$ indexed  by $j$ reduces to

\begin{align} \label{eq:mii_theorem_sidepaths}
&\mathcal{I} \left( \eta^X_{t-\tau}; \eta^Y_t + f(\eta^X_{t-\tau}, \cup_{i} \eta^i_{t-\tau_i}) ; \right.\nonumber\\
&\phantom{\mathcal{I}()}\left\{\eta^j_{t-\tau_j} + f_j(\eta^X_{t-\tau}, \cup_{i\neq j} \eta^i_{t-\tau_i}) \right\}_j  \nonumber\\
&\phantom{\mathcal{I}()}  \left. |~ \mathcal{P}_{Y_t}{\setminus}\mathcal{C}_{X_{t-\tau}{\to}Y_t},  \mathcal{P}(\mathcal{C}_{X_{t-\tau}{\to}Y_t}),\mathcal{N}^{Y_t}_{X_{t-\tau}},\,\mathcal{P}(\mathcal{N}^{Y_t}_{X_{t-\tau}}) \right)\,,
\end{align}

and, if further the ``no contemporaneous dependency''-condition~(\ref{cond:nosidepath_contemp}) holds, to

\begin{align} \label{eq:mii_theorem}
&\mathcal{I} \left( \eta^X_{t-\tau}; \eta^Y_t + f(\eta^X_{t-\tau}, \cup_{i} \eta^i_{t-\tau_i}) ; \right.\nonumber\\
&\left.\phantom{\mathcal{I}()} \left\{\eta^j_{t-\tau_j} + f_j(\eta^X_{t-\tau}, \cup_{i\neq j} \eta^i_{t-\tau_i}) \right\}_j \right)\,,
\end{align}

for linear functions $f,\,f_j$.
\end{Theorem}
The proof is given in Appendix~\ref{sec:app_theorem_mii}.  For the case of a causal triple as shown in Fig.~\ref{fig:interaction_motifs} this further reduces to

\begin{align} \label{eq:mii_theorem_triple}
\mathcal{I} \left( \eta^X_{t-\tau}; \eta^Y_t + (c+ab)\eta^X_{t-\tau} + b\eta^W_{t-\tau_W}; \eta^W_{t-\tau_W} + a\eta^X_{t-\tau} \right)\,,
\end{align}

 from which the special case with Gaussian innovations Eq.~(\ref{eq:interaction_example_chain}) follows.

\section{Discussion} \label{sec:discussion}

%%%%%%%%%%%%%%%%%%%%%%%%%%%%%%%%%%%%%%%%%%%%%%%%%%%%%%%%%%%
\subsection{Relation to linear causal effect theory} \label{sec:causal_effect}

We phrased the idea of causal influence in an information-theoretic setting. Pearl's theory of causal effects \cite{Pearl2000,Pearl2013} can also be embedded in the time series graph framework \cite{Eichler2010a}.
Assuming the time series graph is causally sufficient \cite{Pearl2000} and all dependencies are linear, causal effects can simply by derived from multivariate regressions. Firstly, in analogy to ITY or MIT as a measure of direct link strength, the \emph{path coefficient} of a link is given by the corresponding (typically standardized) coefficient in a multivariate regression of each process on its parents in the time series graph \cite{Wright1934}.
Further, in analogy to ITX or MITP, the linear causal effect of $X_{t-\tau}$ on $Y_t$ also via indirect paths can be estimated by a standardized regression of $Y$ on the multiple regressors $\{X_{t-\tau},\,\mathcal{P}(X_{t-\tau})\}$. The linear \emph{Causal Effect} (CE) \cite{Pearl2000,Pearl2013} is then given by the corresponding (standardized) regression coefficient $r$ belonging to $X_{t-\tau}$,
\begin{align} \label{eq:ce}
{\rm CE}_{X\to Y}(\tau) = r_{Y_t X_{t-\tau}\cdot \mathcal{P}(X_{t-\tau})}\,.
\end{align}
This formulation assumes the ``no contemporaneous dependency''-condition~(\ref{cond:nosidepath_contemp}) for simplicity, but it can be generalized.
The causal effect ${\rm CE}_{X\to Y}(\tau)$ quantifies the change in the expectation of $Y_t$ (in units of its standard deviation) induced by raising the lagged $X_{t-\tau}$ by one standard deviation, while keeping the parents of $X_{t-\tau}$ constant.
Then the total causal effect between lagged processes is simply given by the sum over the product of path coefficients along each causal paths connecting $X_{t-\tau}$ and $Y_t$. For example, for the model~(\ref{eq:xwy_model}) with time series graph in Fig.~\ref{fig:interaction_motifs}(a) the total linear causal effect between $X_{t-2}$ and $Y_t$ is given by $\sqrt{\frac{\Gamma_X}{\Gamma_Y}} (c+ab)$, where the square root contains the normalization by the standard deviations which, however, depends on the autodependency strength and other coefficients here. 
ITX is simply the mutual information with the same conditions as CE (if no neighbors are present), while MITP for this model example (see Eq.~(\ref{eq:mitp_theorem}) or Eq.~(\ref{eq:ana_mitp_sidepath})) is $\frac{1}{2} \ln \left(1+\frac{(c+a b)^2 \sigma_X^2}{b^2 \sigma_W^2+\sigma_Y^2}\right)$. ITX, MITP and CE all depend on the `coupling mechanism' $(c+ab)$, but with different `normalizations'.

Even in linear models, the \emph{Mediated Causal Effect} (MCE) is more difficult to identify \cite{Pearl2000,VanderWeele2015}. 
The causal interpretation is that an indirect effect via the node(s) $W$ measures the increase we would see in $Y_{t}$ while holding $X_{t-\tau}$ and all other intermediate nodes and parents of $X_{t-\tau}$ constant and increasing the node(s) $W$ to whatever value it would obtain under a unit change in $X_{t-\tau}$ while holding the parents of $X_{t-\tau}$ constant \cite{Pearl2000,VanderWeele2015}. 
To identify MCE for the triplet case in model~(\ref{eq:xwy_model}) with time series graph in Fig.~\ref{fig:interaction_motifs}(a) one can subtract from CE the contribution of all paths \emph{not} passing through $W$:
\begin{align} \label{eq:mce}
&{\rm MCE}_{X\to Y|W}(\tau=2) \nonumber\\
&= {\rm CE}_{X\to Y}(2)
                                      - \underbrace{r_{Y_t X_{t-2}\cdot \mathcal{P}(X_{t-2}),W_{t-1},\mathcal{P}(W_{t-1})}}_{
                                       \text{${\rm CE}$ excluding paths through $W$ }} \nonumber\\
&= \sqrt{\tfrac{\Gamma_X}{\Gamma_Y}} a b\,.
                              % &= \sqrt{\tfrac{\Gamma_X}{\Gamma_Y}} a b\,.
\end{align}

Note the additional condition on the parents of $W$ here needed to exclude a confounding of the mediating link from $W$ to $Y$ from the past due to $W_{t-2}$.
This is also the idea behind the interaction information MII which is conditioned on the parents of all intermediate processes to exclude possible confounding. MII is given also by a difference, but of CMIs instead of regressions: $\frac{1}{2} \ln \left(1+\frac{(c+a b)^2 \sigma_X^2}{b^2 \sigma_W^2+\sigma_Y^2}\right) - \frac{1}{2} \ln \left(1+\frac{c^2 \sigma_X^2 \sigma_W^2}{(\sigma_W^2+a^2\sigma_X^2)\sigma_Y^2}\right)$,
% \begin{align}
% & I \left( \eta^X; \eta^Y + (c+ab)\eta^X+b\eta^W \right) - \underbrace{I \left( \eta^X; \eta^Y + (c+ab)\eta^X+b\eta^W ~|~ \eta^W + a\eta^X \right)}_{I \left( \eta^X; \eta^Y + c\eta^X ~|~ \eta^W + a\eta^X \right)}\,,
% \end{align}
where the latter term information-theoretically quantifies the strength of the direct link with coefficient $c$.
The linear framework allows for quantifying the relative influence of paths between two processes by the  `locally' estimated weights making it easy to interpret, but it rests on a linear assumption. Another advantage of the linear approach is that total and indirect effects can also be investigated in the frequency domain in the framework of  \emph{directed transfer functions} \cite{Kaminski2001,Korzeniewska2003,Blinowska2006}.
To some extent causal effects can also be estimated for more general nonlinear structural equation models \cite{spirtes1998using,Pearl2000}, but especially mediated effects are difficult to identify if no strong assumptions are fulfilled \cite{VanderWeele2015}.

\subsection{Advantages and limitations of coupling strength autonomy}

MIT, MITP and MII somewhat disentangle the coupling structure, which is exactly the coupling strength autonomy that makes these measures well-interpretable as measures that solely depend on the ``coupling mechanism'' between $X_{t-\tau}$ and $Y_t$ (and possibly intermediate processes) as shown in the previous sections, autonomous of other external processes. One such possible misleading input ``filtered out'' is autocorrelation, or, more generally, autodependency as has been shown in the model examples. This interpretability is facilitated by the careful conditioning on all possible confounding processes which can be determined from the time series graph (assuming the graph entails all relevant processes, i.e., causal sufficiency \cite{Pearl2000}). In a way, coupling strength autonomy is an information-theoretic description similar to the identifiability of causal effects in Pearl's framework, but this connection needs to be further investigated.

However, the assumptions allowing for such an interpretability are quite restrictive: While arbitrary additive functional dependencies of the interaction processes on external drivers can be conditioned out, the whole interaction mechanism from $X$ to $Y$ via intermediate processes needs to be linear. Note that this does not imply that linear measures can be used instead, because these would not exclude arbitrary nonlinear external drivers.
A further complication is that the potentially high dimensionality due to many external drivers leads to a strong bias in MITP and MII for smaller sample sizes, even for the most advanced information-theoretic estimators employed here \cite{Kraskov2004a,Frenzel2007}. These limitations hamper the added value in interpretability of MITP and MII compared to ITX / IIX. But if no detailed knowledge of the dynamical equations are given, this approach at least is rigorously based on the time series graph encoding the Markovian conditional independence structure as an abstraction of the dynamics. 
Also if the equations are known, but feature highly complex chaotic behavior like toy models from nonlinear dynamics, plots of the measures introduced here like in Figs.~\ref{fig:interaction_motifs},\,\ref{fig:interaction_motifs_synergetic} can help to better understand information transfer in dynamical interactions.

\subsection{Information transfer and complex network theory}
\label{sec:information_efficiency}

In the literature of neuroscience \cite{Bullmore2009,Blinowska2013,Simpson2013} and recently also in climate research \cite{Donges2009,Donges2012thesis}, multivariate datasets are often analyzed using pairwise association measures combined with complex network theory \cite{Newmanbook2010}. Networks are typically reconstructed by thresholding the association matrix (either by some predefined threshold or such that a fixed link density is obtained).  
In interpreting such networks, it is important to take into account the aspect that the network comes from only pairwise associations. For example, the basic principle of transitivity of correlation leads to a lot of spurious links strongly affecting network measures such as the average path length. Typically, short-path lengths in these networks are related to the global efficiency of information transfer, e.g., in the brain \cite{Bullmore2009}, but also in climate \cite{Tsonis2008}. 
But the authors in Refs.~\cite{Bialonski2010,Hlinka2012} have shown that even for a set of entirely independent processes a small world topology (i.e., small average path length and high clustering of the network) emerges. Further, the robustness of a system to random error or perturbations is typically associated with a high clustering coefficient. Also this measure can lead to false interpretations if causality is not taken into account: For example, for the true causal relations $X\to Y \to Z$, there are significant correlations between all pairs and the clustering coefficient of the non-causal network would be maximal. In this simple example an `attack' on node $Y$ in the center certainly disrupts the causal network most because it also destroys the interlink between $X$ and $Z$. But this is not taken into account if the non-causal network is analyzed. In recent years some studies in neuroscience have also applied linear Granger causality methods \cite{Liao2011,Deshpande2011} and bivariate transfer entropy has been applied to climate time series \cite{Hlinka2013}.
%Of course, also the degree and other node-based measures are very different if not the pairwise, but the causal matrix of associations is used to construct the network.

With the measures ITX / MITP and IIX / MII, one can make an attempt to put the notion of shortest paths in an information-theoretic perspective. Instead of counting shortest paths between $X$ and $Y$, ITX or MITP give an appropriate measure of how much information is actually transferred. The interaction informations IIX or MII can then be seen as an alternative to \emph{betweenness centrality} \cite{Freeman1979,Newmanbook2010} originally defined as
\begin{align}
\mathcal{B}(k) = \sum_{i\neq k \neq j} \frac{n_{sp}(k)}{n_{sp}},
\end{align}
where $n_{sp}$ is the total number of shortest paths from node $i$ to node $j$ and $n_{sp}(k)$ is the number of those paths that pass through $k$. In analogy, one can define an aggregated IIX node measure, \emph{causal interaction betweenness} (CIB), as
\begin{align} \label{eq:def_cib}
\mathcal{I}^{\rm CIB}(k) &= \frac{1}{|\mathcal{C}_{k}|} \sum_{(i,j,\tau) \in \mathcal{C}_{k}} |\mathcal{I}^{\rm IIX}_{i{\to}j|\mathbf{k}}(\tau)|\,,
\end{align} 
where $\mathcal{C}_{k}$ is the set of interactions between all non-identical pairs of processes $(i,j)$ at all lags $0<\tau \leq \tau_{\max}$ where $k\neq i,j$ is an intermediate process (at any lags) and $|\mathcal{C}_{k}|$ denotes its cardinality. Here we take the absolute value $|\mathcal{I}^{\rm IIX}_{i{\to}j|\mathbf{k}}(\tau)|$, but one could further distinguish between mediating (positive interaction information) and counteracting (negative interaction information) effects. A linear application of such an approach is discussed in Ref.~\cite{Runge2014c}. 
Instead of IIX, also MII can be used to exclude further biasing confounders at the price of a much higher estimation dimension. Note that $|\mathcal{I}^{\rm IIX}_{i{\to}j|\mathbf{k}}(\tau)|$ does \emph{not} denote a fraction like $ \frac{n_{sp}(k)}{n_{sp}}$ and a more analogous measure to betweenness centrality would be obtained by normalizing each summand in Eq.~(\ref{eq:def_cib}) by the corresponding ITX or MITP,
\begin{align}
\bar{\mathcal{I}}^{\rm CIB}(k) &= \frac{1}{|\mathcal{C}_{k}|} \sum_{(i,j,\tau) \in \mathcal{C}_{k}} \frac{|\mathcal{I}^{\rm IIX}_{i{\to}j|\mathbf{k}}(\tau)|}{I^{\rm ITX}_{i{\to}j}(\tau)}\,,
\end{align} 
which is, however, not robust to outliers for small ITX. 
% Also the related measure ITP can be used in such a way if not the influence of momentary perturbations are of interest, but \emph{all} information entering the system in $X$.
% In our applications to climate data in Sect.~\ref{sec:climate} we will demonstrate how these measures can be used to determine processes important for distributing and mediating information.

\section{Application to climatological time series} \label{sec:climate}
\begin{figure*}[!t]
\begin{center}
\includegraphics[width=1.7\columnwidth]{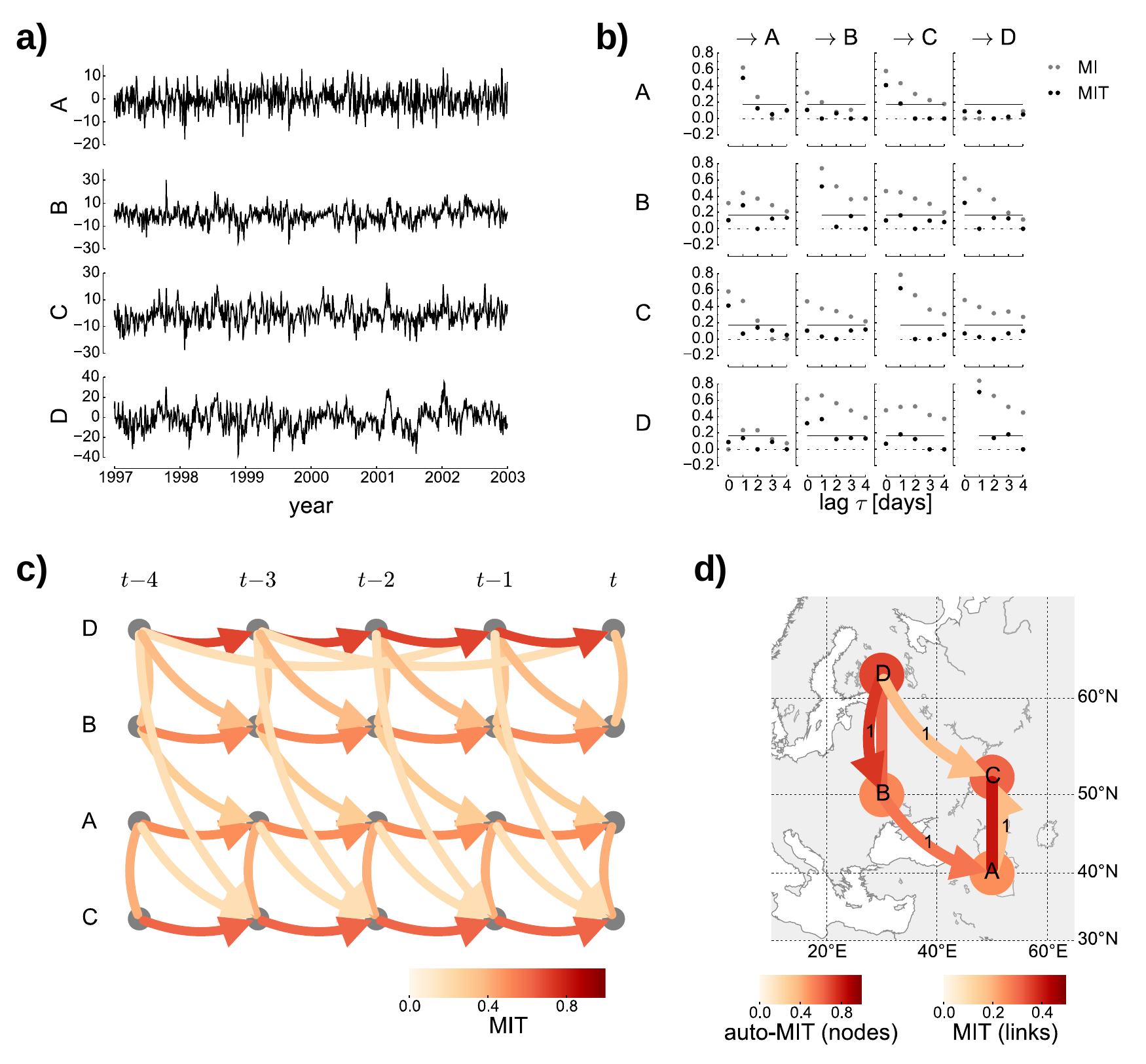}
\end{center}
\caption[Causal analysis of sea level pressure time series in Europe.]{(Color online) Analysis of daily time series of mean sea level pressure with $T=1268$ days. The algorithm to estimate the parents and neighbors was run as in Ref.~\cite{Runge2012prl} using a threshold $I^*=0.015$ $nats$, $\tau_{\max}=4$ days and the CMI nearest-neighbor parameter $k=100$ (larger $k$ have smaller variance which is important for independence tests). 
(a) Anomaly time series (days in winter months November to April only) of the four variables, all units are in hectopascal (hPa) relative to the seasonal mean.
(b) Lag functions of MI and multivariate MIT, here a parameter $k=10$ was chosen to reduce the bias. Also contemporaneous MITs as defined in Ref.~\cite{Runge2012b} are shown. All (C)MI values have been rescaled to the (partial) correlation scale via $I \to\sqrt{1-e^{-2I}}\in [0,1]$ \cite{Cover2006}. The solid lines denote the fixed threshold $I^*=0.015$ (rescaled), which is used to define the time series graph for the path-analysis.
(c) shows the time series graph with the edge color denoting the rescaled MIT strength. Note the different order of the variables to better visualize the causal paths. Repetitions of links emanating from times further than $t-4$ in the past are omitted.
(d) Aggregated visualization as process graph (labels denote the lags and edge and node colors correspond to cross-MIT and auto-MIT, respectively, at the lag with maximum value).
}%entropy $H(Y| \mathcal{P}_{Y} {\setminus}\{X\},\mathcal{P}_{X} )$ 
\label{fig:climate}
\end{figure*}
%%%%%%%%%%%%%%%%%%%%%%%%%%%%%
\begin{table*}[!t]
\centering
\begin{tabular}{l|c|c|c|c}  
\hline
\hline
% Causal path &  ITX  & IIX & MITP & MII  \\ \hline  $W_{t-2}\to\cdots\to X_{t}$ &  $0.19 \pm 0.03$ & & $0.18 \pm 0.02$  & \\  ~~~via $Y_{t-1}$ & & $0.20 \pm 0.03$ & & $0.16 \pm 0.02$  \\ \hline 

%  $W_{t-2}\to\cdots\to Z_{t}$ &  $0.31 \pm 0.02$ & & $0.28 \pm 0.02$  & \\  ~~~via $W_{t-1}$ & & $0.27 \pm 0.02$ & & $0.24 \pm 0.02$  \\ ~~~via $Z_{t-1}$ & & $0.16 \pm 0.02$ & & $0.20 \pm 0.02$  \\ \hline
% $W_{t-3}\to\cdots\to Z_{t}$ &  $0.26 \pm 0.03$ & & $0.24 \pm 0.02$  & \\  ~~~via ($W_{t-2}$, $W_{t-1}$)& & $0.24 \pm 0.03$ & & $0.23 \pm 0.02$  \\ ~~~via $Y_{t-2}$ & & $0.19 \pm 0.03$ & & $0.16 \pm 0.02$  \\ ~~~via $X_{t-1}$ & & $0.17 \pm 0.03$ & & $0.13 \pm 0.02$  \\ ~~~via ($Z_{t-2}$, $Z_{t-1}$)& & $0.25 \pm 0.03$ & & $0.21 \pm 0.02$  \\ \hline
% \end{tabular}
Causal path &  ITX  & IIX & MITP & MII  \\ 
\hline  
$D_{t-2}\to\cdots\to A_{t}$ &  $0.09 \pm 0.06$ & & $0.15 \pm 0.02$  & \\  ~~~via $B_{t-1}$ & & $0.00 \pm 0.06$ & & $0.14 \pm 0.02$  \\
\hline
$D_{t-2}\to\cdots\to C_{t}$ &  $0.26 \pm 0.02$ & & $0.24 \pm 0.02$  & \\  ~~~via $D_{t-1}$ & & $0.22 \pm 0.02$ & & $0.23 \pm 0.02$  \\ ~~~via $C_{t-1}$ & & $0.16 \pm 0.02$ & & $0.18 \pm 0.02$  \\
\hline
$D_{t-3}\to\cdots\to C_{t}$ &  $0.25 \pm 0.02$ & & $0.22 \pm 0.02$  & \\  ~~~via ($D_{t-2}$, $D_{t-1}$)& & $0.22 \pm 0.02$ & & $0.21 \pm 0.02$  \\ ~~~via $B_{t-2}$ & & $0.15 \pm 0.02$ & & $0.13 \pm 0.02$  \\ ~~~via $A_{t-1}$ & & $0.09 \pm 0.02$ & & $0.12 \pm 0.02$  \\ ~~~via ($C_{t-2}$, $C_{t-1}$)& & $0.22 \pm 0.02$ & & $0.20 \pm 0.02$  \\ 
\hline 
\end{tabular}
\caption{Measures of information transfer along selected causal paths for the climatological example of Fig.~\ref{fig:climate}. All (C)MI values have been rescaled to the (partial) correlation scale via $I \to\sqrt{1-e^{-2I}}\in [0,1]$ \cite{Cover2006}. The estimation parameter $k=10$ was chosen as a compromise between low bias and not too high variance, the 68\% confidence interval is based on a bootstrap with $1000$ samples.}
\label{tab:climate}
\end{table*}

To illustrate the causal pathway analysis also on real data, we analyze a climatological dataset of daily mean sea level pressure anomalies (time series with the seasonal cycle removed) in the winter months (November to April) of 1997--2003 \cite{Al2006} at four locations in Eastern Europe indicated as A, B, C, D on the map in Fig.~\ref{fig:climate}(d) which was also analyzed in  \cite{Runge2012prl}. Figure~\ref{fig:climate}(a) depicts the time series. We find that our novel approach of determining not only the information transfer between two processes as in previous work, but also quantifying the exact causal information pathway is especially helpful here and reveals the circular dynamics of the atmospheric processes in this region.

The reconstruction of the causal links with the PC-algorithm was discussed in Ref.~\cite{Runge2012prl}, here we use it in a two-step approach. First, we estimate the preliminary parents and neighbors of all four variables with the causal algorithm as in Ref.~\cite{Runge2012prl} using a fixed significance threshold $I^*=0.015$ $nats$. 
% These are $\mathcal{P}_{A} = \{A_{t-1},\, B_{t-1}\}$, $\mathcal{P}_{B} = \{B_{t-1},\, D_{t-1}\}$, $\mathcal{P}_{C} = \{C_{t-1},\, D_{t-1}\}$, and $\mathcal{P}_{D} = \{D_{t-1}\}$.
These are $\tilde{\mathcal{P}}_{A} = \{A_{t-1},\, B_{t-1}\}$, $\tilde{\mathcal{N}}_{A} = \{C_{t}\}$, $\tilde{\mathcal{P}}_{B} = \{B_{t-1},\, D_{t-1}\}$, $\tilde{\mathcal{N}}_{B} = \{D_{t}\}$, $\tilde{\mathcal{P}}_{C} = \{C_{t-1},\, D_{t-1}\}$, $\tilde{\mathcal{N}}_{C} = \{A_{t}\}$, $\tilde{\mathcal{P}}_{D} = \{D_{t-1}\}$, and $\tilde{\mathcal{N}}_{D} = \{B_{t}\}$.
Secondly, we use these parents and neighbors to estimate MIT values for all links which are plotted in Fig.~\ref{fig:climate}(b) next to MI. Also contemporaneous MIT values using also neighbors as a condition as defined in Ref.~\cite{Runge2012b} are shown. MIT values above the same fixed significance threshold $I^*=0.015$ $nats$ are now considered as the causal links (directed and contemporaneous for $\tau=0$) defining the time series graph shown in Fig.~\ref{fig:climate}(c). We checked that contemporaneous links do not disappear if the contemporaneous neighbors are excluded from the condition in MIT (corresponding to dashed links in Def.~\ref{eq:def_graph_contemp_dashed}). From this graph one can now read off the parents $\mathcal{P}$ and neighbors $\mathcal{N}$ used in the path-based information transfer measures.
This graph also helps to understand why MI has strongly significant values in Fig.~\ref{fig:climate}(b) where MIT is zero. For example, the MI values in panel $C\to D$ can well be explained by past values of $D$, e.g., $D_{t-2}$ acting as a common driver via $D_{t} \leftarrow D_{t-1} \leftarrow D_{t-2}\to C_{t-1}$. 

In the following, we conduct a causal path analysis for the influence of $D$ on $A$ and $C$ at different lags. There are significant ITX values at two and three days lag. From the time series graph (Fig.~\ref{fig:climate}(c)) we can read off the causal paths contributing to the ITX values. In Tab.~\ref{tab:climate} we list the results of an analysis for three causal path interactions.
The interaction $D_{t-2}\to\cdots\to A_{t}$ has only one causal path via $B_{t-1}$, but also contemporaneous sidepaths $D_{t-2}\,{\--} B_{t-2}\to\cdots\to A_{t}$. Here ITX and IIX gave very noisy results (large confidence bounds). MITP, on the other hand, is larger than ITX (as expected from Theorem~\ref{thm:inequality_mitp}) with a much smaller confidence interval. Here MII via $B_{t-1}$ explains all of the MITP within error bounds as expected since it is the only intermediate node and no direct link exists.
% Note that here the contemporaneous path $D_{t-2}\,\text{- - -}\,B_{t-2}\to A_{t-1}\to A_{t}$, even though not a causal mechanism, also contributes to ITX and MITP.
Next, we turn to the more interesting influence of $D$ on $C$. At a lag of two days MITP is slightly smaller than ITX, which, as discussed in Sect.~\ref{sec:sidepaths}, is due to finite sample bias. 
The indirectness of the interaction $D_{t-2}\to\cdots\to C_{t}$ here stems from the two paths $D_{t-2}\to D_{t-1}\to C_{t}$ and $D_{t-2}\to C_{t-1}\to C_{t}$ via autodependencies (Fig.~\ref{fig:climate}(c)). The interaction analyses with IIX and MII here both indicate that a slightly larger part of the ITX is mediated via $D_{t-1}$ rather than $C_{t-1}$ (Tab.~\ref{tab:climate}) in line with the higher auto-MIT strength of the autodependency within $D$.
At a lag of three days the interaction $D_{t-3}\to\cdots\to C_{t}$ has many more paths not only via autodependencies, but also via $B_{t-2}$ and $A_{t-1}$ (and also non-causal contemporaneous sidepaths). While also here the auto-dependencies together with the direct link $D_{t-1}\to C_t$ strongly contribute to ITX (Tab.~\ref{tab:climate}), the path $D_{t-3}\to B_{t-2}\to A_{t-1}\to C_{t}$ seems to be relevant, too, as indicated by the significant IIX and MII values through these nodes.

This causal picture of a counter-clockwise `flow of entropy' is consistent with the dynamical processes governing the lower and middle atmosphere circulation in the considered area. One usually observes a superposition of westerly winds with traveling extratropical counter-clockwise cyclones that traverse the area and whose trajectories are regulated by the aforementioned westerlies \cite{Palmen1969}. Consistent with the causal lags of one or two days, these processes act on short daily time scales.
Note that the variables were defined in an ad-hoc manner by the locations of grid points here, but one can better isolate subprocesses of complex systems by a suitable dimension reduction, see \cite{Vejmelka2014,Runge2014c} for an application to the global atmospheric pressure system.

\section{Conclusions} \label{sec:conclusions}

This work expanded the approach introduced in Ref.~\cite{Runge2012b} which considered information-theoretic measures to quantify the strength of links in causal time series graphs. Here the goal was to quantify indirect causal interactions and how much intermediate processes mediate or counteract an interaction. Our approach is more focused on a detailed picture of an interaction mechanism between two variables and complements concepts aimed at decomposing predictive information about a target variable $Y$.

The two considered pairs of measures ITX / IIX and MITP / MII for a causal interaction $X_{t-\tau}\to\cdots\to Y_t$ have in common the idea to extract information originating in process $X$ only at the lagged time $t-\tau$ and are conditioned in order to measure only information transfer along causal paths. MITP further attempts to exclude the influence of other drivers of $Y$ or intermediate path nodes by conditioning out the parents of all processes involved in the causal interaction. As a further step, IIX and MII quantify the mediating or counteracting effect of intermediate processes on causal paths to an interaction mechanism to determine the relative importance of pathways of causal information transfer. 
In extensions of the coupling strength autonomy theorem \cite{Runge2012b}, for certain model classes MITP and MII allow to entirely isolate the quantification of the interaction mechanism from other driving mechanisms. Then the values of MITP and MII can be solely related to the coefficients belonging to the indirect interaction mechanism between $X$ and $Y$ making them well interpretable not only information-theoretically, but also relating their value to the underlying dynamics.

Generally, however, the value of MIT or MITP remains hard to interpret for nonlinearly intertwined complex systems, but their information-theoretic definition and foundation based on the Markov structure of the process allows to quantify a rigorous notion of causal information transfer as an abstraction of the dynamics. The novel measures can also be helpful in understanding dynamical interactions in toy models from nonlinear dynamics.
While the absolute values of ITX and MITP, measured in $nats$, cannot be simply related to units of the variables like linear measures, the values of the interaction measures IIX and MII can be used to quantify how much of the information transfer can be attributed to individual intermediate processes.
The goal of information-theoretic measures is not a complete understanding of the dynamics of the system which can only be achieved by experiments or detailed modeling. Then causal effect quantifiers such as proposed in Pearl \cite{Pearl2000} or \cite{Smirnov2014} are good starting points. 

The climatological analysis underlines the importance of inferring mechanism delays and pathways for physical interpretations and serves as a first step to study more complex systems in climate and beyond.
More exploratory studies in the spirit of functional network analysis, but with a rigorous definition of information transfer, can be based on the aggregate measures introduced in Sect.~\ref{sec:information_efficiency}. A linear application of such an approach is demonstrated in Ref.~\cite{Runge2014c}. 
As a further outlook, it will be an interesting avenue of research to connect the time series graph-based framework of information transfer to recent concepts of \emph{synergistic information sharing} \cite{Olbrich2015,Barrett2015}. In Ref.~\cite{Runge2014b} synergistic effects are studied with respect to optimal prediction schemes.

%%%%%%%%%%%%%%%%%%%%%%%%%%%%%%%%%%%%%%%%%%

\section*{Acknowledgments}

The early stages of this work benefited from discussions with Bernd Pompe and Jobst Heitzig. Vladimir Petoukhov helped in interpreting the climatological example. 
This work was supported by the German National Academic Foundation, a Humboldt University Postdoctoral Fellowship and the German Federal Ministry of Science and Education (Young Investigators Group CoSy-CC$^2$, grant no. 01LN1306A). 
A \emph{Python} script to estimate the causal network can be obtained from the author's website at \texttt{www.pik-potsdam.de/members/jakrunge}.

The author declares no conflict of interest.

\appendix

% \section{Forward-selection as a causal reconstruction algorithm} \label{sec:forward-selection-failure}

% A forward-selection algorithm alone cannot be used to reconstruct graphical models. The non-uniform embedding vector reconstructed with the scheme proposed in Refs.~\cite{Faes2011,Kugiumtzis2013a,Faes2015a}, therefore, can contain also non-causal drivers: In the first step of forward selection, the variable which maximizes the MI with $Y$ is picked. Then causal drivers are iteratively determined based on how much information they contain \emph{additionally} to the already chosen variables using CMI. A simple example demonstrating why this can select spurious drivers is

% \begin{align}
% X_t &= a (Z^1_{t-1} + Z^2_{t-1}) + \eta^X_t \nonumber\\
% Y_t &= c (Z^1_{t-2} + Z^2_{t-2}) + \eta^Y_t\,,
% \end{align}

% with Gaussian unit variance white noise processes $Z^{1,2},\,\eta^{X,Y}$. Clearly, here $Z^1$ and $Z^2$ are the causal driving variables of $Y$ at lag $t-1$ in the sense of Pearl, i.e., an intervention in $Z^1$ or $Z^2$ changes the distribution of $Y$ while an intervention in $X$ does not. But for positive $c$ and $a>\frac{1}{\sqrt{2}}$ the MI between $X_{t-1}$ and $Y_t$ is larger than the MI between any of the $Z^{1,2}_{t-2}$ and $Y_t$. Thus, a non-causal variable is selected first which unnecessarily increases the dimension of the non-uniform embedding vector. By testing each variable again conditional on the remaining variables in the vector (backward-elimination), spurious drivers can finally be removed \cite{Kugiumtzis2013a,Faes2015a,Sun2014a,Sun2015}. 

\section{Proofs of theorems}

\subsection{Proof of Inequality Theorem~\ref{thm:inequality_mitp}} \label{thm:inequality_mitp_proof}

The Inequality Theorem~\ref{thm:inequality_mitp} can be proven similarly to the inequalities among ITY and MIT in Ref.~\cite{Runge2012b}. 
To simplify notation, we drop the time indices and write $X$ for $X_{t-\tau}$, $Y$ for $Y_{t}$, $\mathcal{N}^{Y}_{X}$ for $\mathcal{N}^{Y_t}_{X_{t-\tau}}$, and  $\mathcal{C}_{X\to Y}$ for  $\mathcal{C}_{X_{t-\tau}\to Y_t}$.
% \mathcal{P}_{Y}{\setminus}\mathcal{C}_{X\to Y},  \mathcal{P}(\mathcal{C}_{X{\to}Y})

\begin{proof}
We define $\tilde{\mathcal{P}}$ to be the set of parents of both $Y$ and the path nodes $\mathcal{C}_{X\to Y}$ (including $X$) that is not already included in the conditions of ITX $(\mathcal{P}_X,\mathcal{N}^{Y}_{X},\,\mathcal{P}(\mathcal{N}^{Y}_{X})$), i.e., $\tilde{\mathcal{P}}=(\mathcal{P}_{Y}{\setminus}\mathcal{C}_{X\to Y},  \mathcal{P}(\mathcal{C}_{X{\to}Y})) \setminus (\mathcal{P}_{X},\mathcal{N}^{Y}_{X},\,\mathcal{P}(\mathcal{N}^{Y}_{X}))$. 
Then it generally holds that $I(X;\tilde{\mathcal{P}}~|~\mathcal{P}_{X},\mathcal{N}^{Y}_{X},\,\mathcal{P}(\mathcal{N}^{Y}_{X}))=0$: Firstly, all paths arriving at $X$ from the past are surely blocked (see Sect.~\ref{sec:tsg_paths}) by $\mathcal{P}_{X}$ because they contain the motifs $\to \blacksquare \to X$ or ${\--}\,\blacksquare \to X$ which are both blocked. Further, also contemporaneous sidepaths are blocked by $(\mathcal{N}^{Y}_{X},\,\mathcal{P}(\mathcal{N}^{Y}_{X}))$ and there are also no directed causal paths from $X$ to any node in $\tilde{\mathcal{P}}$ since, by definition, such a node would belong to $\mathcal{C}_{X\to Y}$. 
We now apply the chain rule on the (multivariate) CMI $I(X;(Y,\tilde{\mathcal{P}})| \mathcal{P}_{X},\mathcal{N}^{Y}_{X},\,\mathcal{P}(\mathcal{N}^{Y}_{X}))$ twice:

\begin{align}
& I(X; (Y,\tilde{\mathcal{P}})| \mathcal{P}_{X},\mathcal{N}^{Y}_{X},\,\mathcal{P}(\mathcal{N}^{Y}_{X})) \nonumber \\ 
&= I(X;Y |  \mathcal{P}_{X},\mathcal{N}^{Y}_{X},\,\mathcal{P}(\mathcal{N}^{Y}_{X})) \nonumber\\
&\phantom{=}+ 
\underbrace{ I(X;\tilde{\mathcal{P}}| \mathcal{P}_{X},\mathcal{N}^{Y}_{X},\,\mathcal{P}(\mathcal{N}^{Y}_{X}), Y) }_{\geq 0} \\
& = \underbrace{I(X; \tilde{\mathcal{P}}|\mathcal{P}_{X},\mathcal{N}^{Y}_{X},\,\mathcal{P}(\mathcal{N}^{Y}_{X}))}_{= 0} \nonumber \\
&\phantom{=}+ I(X;Y| \tilde{\mathcal{P}},\mathcal{P}_{X},\mathcal{N}^{Y}_{X},\,\mathcal{P}(\mathcal{N}^{Y}_{X})) \\
& \implies  I(X;Y|\tilde{\mathcal{P}},\mathcal{P}_{X},\mathcal{N}^{Y}_{X},\,\mathcal{P}(\mathcal{N}^{Y}_{X})) \nonumber\\
&\phantom{\implies} = I(X;Y|\mathcal{P}_{Y}{\setminus}\mathcal{C}_{X\to Y},  \mathcal{P}(\mathcal{C}_{X{\to}Y}),\mathcal{N}^{Y}_{X},\,\mathcal{P}(\mathcal{N}^{Y}_{X})) \nonumber\\
&\phantom{\implies} \geq I(X;Y|\mathcal{P}_{X},\mathcal{N}^{Y}_{X},\,\mathcal{P}(\mathcal{N}^{Y}_{X}))\,.
\end{align}
\end{proof}

\subsection{Further information-theoretic properties}

% Shannon's entropy $H$ \cite{Shannon1948,Shannon1963} is a measure of the uncertainty about outcomes of a process.
 % for a real-valued multivariate process $\mathbf{X}=(X_1,\,X_2,\,\ldots,X_N)$ it is given by
% \begin{align}
% &H(X_1,\ldots,X_N) = \nonumber\\
% &-\int \cdots \int p(x_1,\ldots,x_N) \ln p(x_1,\ldots,x_N) \,dx_1\cdots dx_N\,,
% \end{align}
% where $p(\cdot)$ denotes the probability density function. Here we use the natural logarithm to measure entropy and derived measures in $nats$.
% \emph{Mutual information} (MI), on the other hand, is a measure of the reduction of this uncertainty if another process is measured. The Shannon type MI between two real-valued (possibly multivariate) processes $X$ and $Y$ is defined as

% \begin{align} 
% I(X;Y) &= \int \int p(x,y) \ln \frac{p(x,y)}{p(x)p(y)} \,dx dy \label{eq:mi_indep}\\
%        &= H(Y) - H(Y|X) = H(X) - H(X|Y) \label{eq:mi_cond}\,,
% \end{align}

% i.e., in the form of Eq.~(\ref{eq:mi_cond}) as the difference between the uncertainty in $Y$ and the remaining uncertainty if $X$ is already known (and vice versa). Or in Eq.~(\ref{eq:mi_indep}) as a certain Kullback-Leibler distance \cite{Kullback1951,Cover2006} between the distributions $p(x,y)$ and the distribution for the independent case $p(x)p(y)$. 

Some further fundamental properties of information-theoretic quantities are important for the coupling strength autonomy theorems. 
% Further, using Jensen's inequality \cite{Cover2006} one can show that MI is always non-negative (which holds for the continuous as well as the discrete case).
% The random variables $X,\,Y,\,Z$ can also be multivariate and we will sometimes use multivariate CMIs like $I((X,W);Y|Z)$ where the colon always separates the arguments. For multivariate CMIs also a \emph{chain rule} holds which we frequently utilize in this article:
% \begin{align}  \label{eq:chain_rule_cmi}
% I(X_1,\ldots,X_N;Y|Z) &= I(X_1;Y|Z) + I(X_2,\ldots,X_N;Y|X_1,Z) \nonumber \\ 
%                         &\vdots \nonumber \\
%                     &= \sum_{i=1}^N I(X_i;Y|\cup_{j=1}^{i-1} X_j,Z).
% \end{align}
The data processing inequality \cite{Cover2006} states that 
\begin{align} \label{eq:data_processing_ineq}
I(X;f(Y)|Z)\leq I(X;Y|Z)\,,
\end{align}
i.e., manipulating $Y$ (which can also be a vector) by some function $f$ can only reduce the shared information. Note, however, that equality holds for smooth uniquely invertible transformations such as linear rescalings of $X$, $Y$ or $Z$ under which CMI is invariant \cite{Kraskov2004a}.
% This can easily be seen from Eq.~(\ref{eq:entropy_jacobian}), because such transformations alter the joint density to
% \begin{align}
% p_{X'Y'Z'}(x',y',z')=J(x,x')J(y,y')J(z,z')p_{XYZ}(x,y,z)\,,
% \end{align}
% where $J(\cdot,\cdot)$ denotes the Jacobians. Correspondingly, the marginal entropies are altered and in the formula for CMI (Eq.~(\ref{eq:def_cmi})) the second terms in Eq.~(\ref{eq:entropy_jacobian}) then cancel out.
For random variables $Y$ and $W$ and an arbitrary function $f$ we have that
\begin{align}
H(Y + f(W) | W ) &= \int p(w) H(Y + f(W) | W=w ) dw \nonumber\\
            & = \int p(w) H(Y | W=w ) dw \nonumber \\
            &= H(Y|W)\,, \label{eq:trans_inv}
\end{align}
because $f(W)$ for $W=w$ is a fixed constant and entropies are translationally invariant. In particular, $H(f(W) | W )=0$. This property also holds for the joint entropy and with another arbitrary function $g$ it follows for CMI that
\begin{align} \label{eq:cmi_trans_inv}
I(X+g(Z);Y+f(W)|Z,W)= I(X;Y|Z,W).
\end{align}
Also here, $I(X; f(W)| W )=0$.
Last, conditions that are conditionally independent of the joint vector $(X, Y)$ given $Z$ can be dropped:
\begin{align} \label{eq:cmi_drop_conds}
& I\left((X,Y);W~|~Z\right) = 0  \nonumber \\
& \Longrightarrow ~ I(X;Y~|~W,Z) = I(X;Y~|~Z) \,,
\end{align}
which can be derived from the fundamental decomposition and weak union properties of conditional independence relations. This relation also holds without the condition on $Z$.

\subsection{Proof for momentary information transfer along paths}
\label{sec:app_theorem_mitp}

Also here, to simplify notation we drop the time indices and write $X$ for $X_{t-\tau}$, $Y$ for $Y_{t}$, $\mathcal{N}^{Y}_{X}$ for $\mathcal{N}^{Y_t}_{X_{t-\tau}}$, and  $\mathcal{C}_{X\to Y}$ for  $\mathcal{C}_{X_{t-\tau}\to Y_t}$.
In the theorem, we denoted those parents of $Y$ that are in the path nodes $\mathcal{C}_{X\to Y}$ defined in Eq.~(\ref{eq:causal_paths}) as $\mathcal{P}^\mathcal{C}_Y=\mathcal{P}_Y \cap \mathcal{C}_{X\to Y}$ and correspondingly for other path nodes $\mathcal{P}_i^\mathcal{C}$ indexed by $i$. Also note that $X$ is included in the set of path nodes.

\begin{proof}
We insert the dependencies assumed for $X$ and $Y$ in Eq.~(\ref{cond:additivity_paths}) in the definition of MITP (Eq.~(\ref{eq:def_mitp})):
\begin{align}
&I^{\rm MITP}_{X{\to}Y}\nonumber\\
&\overset{\phantom{\text{Eq.~(\ref{eq:def_mitp})}}}{=} I(X;Y~|~  \mathcal{P}_{Y}{\setminus}\mathcal{C}_{X\to Y},  \mathcal{P}(\mathcal{C}_{X{\to}Y}),\mathcal{N}^{Y}_{X},\,\mathcal{P}(\mathcal{N}^{Y}_{X}) )\\
&\overset{\text{Eq.~(\ref{cond:additivity_paths})}}{=} I(g_X(\mathcal{P}_{X}) + \eta^X;f_Y(\mathcal{P}^\mathcal{C}_Y) + g_Y (\mathcal{P}_{Y}\setminus \mathcal{P}^\mathcal{C}_Y) +  \eta^Y \nonumber\\
&\phantom{\overset{\text{Eq.~(\ref{cond:additivity_paths})}}{=} I()}~|~ \mathcal{P}_{Y}{\setminus}\mathcal{C}_{X\to Y},  \mathcal{P}(\mathcal{C}_{X{\to}Y}),\mathcal{N}^{Y}_{X},\,\mathcal{P}(\mathcal{N}^{Y}_{X}) ) \\
&\overset{\text{Eq.~(\ref{eq:cmi_trans_inv})}}{=} I( \eta^X;f_Y(\mathcal{P}^\mathcal{C}_Y)+  \eta^Y \nonumber\\
&\phantom{\overset{\text{Eq.~(\ref{eq:cmi_trans_inv})}}{=} I()} ~|~ \mathcal{P}_{Y}{\setminus}\mathcal{C}_{X\to Y},  \mathcal{P}(\mathcal{C}_{X{\to}Y}),\mathcal{N}^{Y}_{X},\,\mathcal{P}(\mathcal{N}^{Y}_{X}) ).
\end{align}

In the theorem, $f_Y$ is assumed linear and we also assumed all other path nodes $W^{(i)}\in \mathcal{C}_{X{\to}Y}$ to linearly depend on each other by Eq.~(\ref{cond:linearity_path_nodes}), where dependencies on external nodes were only assumed additive. Then,
\begin{align}
I^{\rm MITP}_{X{\to}Y}
&\overset{\text{Eq.~(\ref{eq:cmi_trans_inv})}}{=} I( \eta^X;f(\eta^X, \cup_{i} \eta^i)+  \eta^Y \nonumber\\
&\phantom{\overset{\text{Eq.~(\ref{eq:cmi_trans_inv})}}{=} I()} ~|~ \mathcal{P}_{Y}{\setminus}\mathcal{C}_{X\to Y},  \mathcal{P}(\mathcal{C}_{X{\to}Y}),\mathcal{N}^{Y}_{X},\,\mathcal{P}(\mathcal{N}^{Y}_{X}) ),
\end{align}
for some linear function $f$ yielding Eq.~(\ref{eq:mitp_theorem_sidepaths}). 

Now under the  ``no contemporaneous dependency''-condition~(\ref{cond:nosidepath_contemp}) it holds that $\mathcal{N}^{Y}_{X}=\emptyset$ and further
\begin{align} \label{eq:joint_zero}
I\left( (\eta^X, \eta^Y, \cup_{i} \eta^i)~;~\mathcal{P}_{Y}{\setminus}\mathcal{C}_{X\to Y},  \mathcal{P}(\mathcal{C}_{X{\to}Y})   \right) = 0\,,
\end{align}
which can be derived graph-theoretically exploiting Markov properties as follows: Firstly, since the noise terms $(\eta^X, \eta^Y, \cup_{i} \eta^i)$ of the path nodes in $\mathcal{C}_{X{\to}Y}$ and $Y$ are i.i.d., they are independent of all those processes in $(\mathcal{P}_{Y}{\setminus}\mathcal{C}_{X\to Y},  \mathcal{P}(\mathcal{C}_{X{\to}Y}))$ with paths ending with a directed arrow at any of the path nodes $\mathcal{C}_{X{\to}Y}$ or $Y$. Secondly, by definition of $\mathcal{C}_{X{\to}Y}$ there are no directed paths from any node in $\mathcal{C}_{X{\to}Y}$ toward $(\mathcal{P}_{Y}{\setminus}\mathcal{C}_{X\to Y},  \mathcal{P}(\mathcal{C}_{X{\to}Y}))$. Last, contemporaneous sidepaths from any node in $\mathcal{C}_{X{\to}Y}$ to $(\mathcal{P}_{Y}{\setminus}\mathcal{C}_{X\to Y},  \mathcal{P}(\mathcal{C}_{X{\to}Y}))$ are excluded by the ``no contemporaneous dependency''-condition~(\ref{cond:nosidepath_contemp}).

Further, from Eq.~(\ref{eq:joint_zero}) we find that $I \left( (\eta^X, f(\eta^X, \cup_{i} \eta^i)+  \eta^Y)~;~ \mathcal{P}_{Y}{\setminus}\mathcal{C}_{X\to Y},  \mathcal{P}(\mathcal{C}_{X{\to}Y})\right) = 0$ due to the data processing inequality~(\ref{eq:data_processing_ineq}) and therefore we can drop the conditions due to Eq.~(\ref{eq:cmi_drop_conds}),
\begin{align}
I^{\rm MITP}_{X{\to}Y}
&\overset{\text{Eq.~(\ref{eq:cmi_drop_conds})}}{=} I( \eta^X;f(\eta^X, \cup_{i} \eta^i)+  \eta^Y),
\end{align}

yielding Eq.~(\ref{eq:mitp_theorem}). Note that since the dynamical noise is i.i.d. and $0<\tau_i<\tau$, it holds that $(\eta^X, \eta^Y) ~\ci~ \eta^i ~~\forall~~ i$ and $\eta^X ~\ci~ \eta^Y$.
\end{proof}

This proof also includes the proof for the MIT coupling strength autonomy theorem as a special case, but in a much shorter form than in Ref.~\cite{Runge2012b}: If $\mathcal{C}_{X_{t-\tau}\to Y_t}=\{X_{t-\tau}\}$, and under the ``no sidepath''-constraint in Ref.~\cite{Runge2012b}, the conditions on the neighbors can be dropped and MITP collapses to MIT. Since then also $f(\eta^X_{t-\tau}, \cup_{i} \eta^i_{t-\tau_i})=f(\eta^X_{t-\tau})$, Eq.~(\ref{eq:mitp_theorem})  reduces to the same form as in Ref.~\cite{Runge2012b}.

% since $\mathcal{C}_{X\to Y}$ entails nodes on all directed causal paths emanating from $X_{t-\tau}$ and ending in $Y_t$, also all sidepath nodes are included in this set. This implies that the innovations $\eta^X, \cup_{i} \eta^i$ are independent of the external parents, which can be expressed with a multivariate MI as

% \begin{align}
% I\left((\eta^X,\,\eta^Y,\,\cup_{i} \eta^i)~;~( \mathcal{P}_{Y}{\setminus}\mathcal{C}_{X\to Y},  \mathcal{P}(\mathcal{C}_{X{\to}Y})) \right)=0\,,
% \end{align}
% and we can drop the condition due to Eq.~(\ref{eq:cmi_drop_conds}),

\subsection{Proof for momentary interaction information}
\label{sec:app_theorem_mii}

Using the same assumptions as for Theorem~\ref{thm:autonomy_mitp}, the dependencies of  momentary interaction information between $X$, $Y$ and intermediate processes $\mathbf{W}=(W^{(1)}_{t-\tau_{1}},\,W^{(2)}_{t-\tau_{2}}\,\ldots)\in \mathcal{C}_{X_{t-\tau}\to Y_t}\setminus\{X_{t-\tau}\}$ indexed  by $j$ can be simplified exploiting the same arguments as above.

\begin{proof}

\begin{align}
&\mathcal{I}^{\rm MII}_{X{\to}Y|\mathbf{W}} \nonumber\\
&\overset{\phantom{\text{Eq.~(\ref{eq:cmi_trans_inv})}}}{=}\mathcal{I} \left(X;Y;\mathbf{W}~|~\mathcal{P}_{Y}{\setminus}\mathcal{C}_{X,Y},  \mathcal{P}(\mathcal{C}_{X{\to}Y}),\mathcal{N}^{Y}_{X},\,\mathcal{P}(\mathcal{N}^{Y}_{X}) \right) \\
&\overset{\text{Eq.~(\ref{eq:cmi_trans_inv})}}{=}\mathcal{I} ( \eta^X;f(\eta^X, \cup_{i} \eta^i)+  \eta^Y; \left\{\eta^j + f_j(\eta^X, \cup_{i\neq j} \eta^i) \right\}_j  \nonumber\\
&\phantom{\overset{\text{Eq.~(\ref{eq:cmi_trans_inv})}}{=}\mathcal{I} ()} ~|~\mathcal{P}_{Y}{\setminus}\mathcal{C}_{X{\to}Y},  \mathcal{P}(\mathcal{C}_{X{\to}Y}),\mathcal{N}^{Y}_{X},\,\mathcal{P}(\mathcal{N}^{Y}_{X}) ) \\
&\overset{\text{Eq.~(\ref{eq:cmi_drop_conds})}}{=}\mathcal{I}( \eta^X;f(\eta^X, \cup_{i} \eta^i)+  \eta^Y; \left\{\eta^j + f_j(\eta^X, \cup_{i\neq j} \eta^i) \right\}_j )\,,
\end{align}

where the last step is valid only under the  ``no contemporaneous dependency''-condition Eq.~(\ref{cond:nosidepath_contemp}) giving Eq.~(\ref{eq:mii_theorem}) with linear functions $f,\,f_j$.
\end{proof}

% In Ref.~\cite{Runge2014d} these results are further generalized for contemporaneous links.

% \bibliography{extracted.bib}

\bibliography{Runge_manuscript_pre_bibliography.bib}
% \bibliography{/home/jakob/work/library}
\bibliographystyle{myapsrev.bst} %\bibliographystyle{unsrt} %\bibliographystyle{plain}

\end{document}